\documentclass[11pt]{article}

\usepackage{graphicx}
\usepackage{amsmath}
\usepackage{latexsym}
\usepackage{amssymb}
\usepackage{setspace,cite}

\DeclareSymbolFont{AMSb}{U}{msb}{m}{n}
\DeclareMathSymbol{\N}{\mathbin}{AMSb}{"4E}
\DeclareMathSymbol{\Z}{\mathbin}{AMSb}{"5A}
\DeclareMathSymbol{\R}{\mathbin}{AMSb}{"52}
\DeclareMathSymbol{\Q}{\mathbin}{AMSb}{"51}
\DeclareMathSymbol{\I}{\mathbin}{AMSb}{"49}
\DeclareMathSymbol{\C}{\mathbin}{AMSb}{"43}

\newcommand{\rp}{\mathbb{R}\mathbb{P}^3}

\begin{document}

\title{Causal particle detectors and topology}
\author{Paul Langlois\footnote{Electronic Address: pmxppl@nottingham.ac.uk} 
\\ \textit{School of Mathematical Sciences, University of Nottingham} 
\\ 
\textit{Nottingham NG7 2RD, UK}}
\date{\today}
\maketitle

\begin{abstract}

We investigate particle detector responses in some topologically non-trivial
spacetimes. We extend a recently proposed regularization 
of the massless scalar field Wightman function in $4$-dimensional Minkowski 
space to arbitrary dimension, to the massive scalar field, to quotients of Minkowski space under 
discrete isometry groups and to the massless Dirac field. 
We investigate in detail the transition rate of inertial and uniformly accelerated
detectors on the quotient spaces under groups generated by 
$(t,x,y,z)\mapsto(t,x,y,z+2a)$, $(t,x,y,z)\mapsto(t,-x,y,z)$, 
$(t,x,y,z)\mapsto(t,-x,-y,z)$, $(t,x,y,z)\mapsto(t,-x,-y,z+a)$ and some
higher dimensional generalizations. For motions in at constant $y$ and $z$ on the latter three 
spaces the response is time dependent.
We also discuss the response of static detectors on the $\rp$ geon and inertial
detectors on $\rp$ de Sitter space via their associated global embedding
Minkowski spaces (GEMS). The response
on $\rp$ de Sitter space, found both directly and 
in its GEMS, provides support for the validity of applying the GEMS procedure
to detector responses and to quotient spaces such as $\rp$ de Sitter space and
the $\rp$ geon where the embedding spaces are Minkowski spaces with suitable
identifications.

\end{abstract}

\section{Introduction}

In this paper we investigate particle detector models in topologically non-trivial
spacetimes. The aims are two-fold. Firstly we discuss a recent paper by Schlicht
\cite{sc:schlicht} in which the usual regularization procedure for the positive
frequency Wightman function \cite{bd:book} is criticized in the context of particle detectors
in Minkowski space and an alternative is proposed. We extend the regularization introduced 
by Schlicht to Minkowski space of arbitrary dimension, to the massive scalar field, to automorphic fields and to
the massless Dirac field. Secondly we present detector responses on certain
topologically non-trivial spacetimes, with the aim of investigating the
effect of the non-trivial topology on the Unruh and Hawking effects 
\cite{u:unruh,h:rad}.

Particle detector models in the context of quantum field 
theory in curved spacetime were first considered by Unruh
\cite{u:unruh}, who proved his now famous result that 
a uniformly accelerated detector in Minkowski space responds
as in a thermal bath at the temperature $T=a/(2\pi)$.
Unruh's model consists of a particle in a box coupled to the quantum field.
The detection of a particle is indicated by the excitation of the detector by
the field.
Shortly after Unruh, DeWitt \cite{d:dewitt} considered a simpler 
model consisting of a monopole moment operator coupled to the field
along the detector's worldline. This is the model subsequently considered
most in the literature. Good reviews of the early literature are found in
\cite{bd:book,tk:takagi}.

In the usual derivation of the Unruh effect using DeWitt's monopole 
detector the detection is considered in the asymptotic regions where the detector
is switched on in the infinite past and off in the infinite future.
Recently Schlicht \cite{sc:schlicht} has considered the case of a detector 
switched on in the infinite past but read at a finite time $\tau$ and has highlighted the
importance of choosing a suitable regularization of the Wightman function.
In particular Schlicht shows that a naive $i\epsilon$-prescription, as is usually used
in the asymptotic case \cite{bd:book}, for a uniformly accelerated detector, leads
to a $\tau$-dependent and thus presumably unphysical response.
Schlicht proposes an alternative regularization procedure by considering the
monopole detector as the limit of a rigid detector with spatial extension. The monopole moment
operator is coupled to the field via a smeared field operator where the smearing is performed in the
detector's rest frame.
He recovers the usual responses in the case of inertial and uniformly accelerated motion. He further
considers a trajectory which smoothly interpolates between the two obtaining physically
reasonable results.

We begin here in section \ref{sec-linmink} by extending the regularization introduced 
by Schlicht \cite{sc:schlicht} to a massless scalar field in Minkowski space of arbitrary dimension.
We recover the expected responses for inertial and uniformly accelerated detectors.
In section \ref{sec:massive} we extend the regularization further to a massive scalar field
in four-dimensional Minkowski space.

In section \ref{sec:autodetector} we consider detectors on quotient spaces of Minkowski 
space under certain discrete isometry groups, via detectors coupled to automorphic fields
\cite{bD:auto,bd:em} in Minkowski space. We begin by extending Schlicht's regularization 
to these quotient spaces. 
We then present a number of responses on a number of specific spacetimes and trajectories of interest.
Our main motivation is to investigate the effect of non-trivial topology on the
Hawking and Unruh effects. In particular our interest lies in the (non)-thermality
of the Hartle-Hawking-like vacuum on the $\rp$ geon black hole spacetime as seen
by static observers \cite{lm:geon,pl:langlois} and the Euclidean-like vacuum state
on $\rp$ de Sitter space \cite{jk:desit}.
We consider inertial and uniformly accelerated observers on the quotient spaces of Minkowski space
under the involutions $J_0:(t,x,y,z)\mapsto(t,x,y,z+2a)$ and $J_-:(t,x,y,z)\mapsto(t,-x,-y,z+a)$ (denoted here by $M_0$ and 
$M_-$ respectively), in the Minkowski-like vacuum states.
$M_0$ and $M_-$ may be used to model the Hawking effect on the Kruskal manifold
and the $\rp$ geon respectively \cite{lm:geon,pl:langlois} and to illustrate the
affect of the topology on the Unruh effect in flat spacetimes.
We also consider inertial and uniformly accelerated detectors on Minkowski
space with an infinite flat plane boundary, on a conical spacetime and
on some higher dimensional generalizations.

For motions perpendicular to the boundary the boundary spacetime serves as a 
simpler model in which many of the features of the responses on $M_-$ are present.
Further the responses are interesting in their own right. The boundary spacetime
with Dirichlet boundary conditions has been used to investigate the detection of
negative energy densities \cite{do:daviesott} and may have relevance to the
quantum inequalities program \cite{fo:ford,fo:ford2,fo:phenford,fo:phenford2,fo:phenfordrom,f:fewster,f:fewster2,f:fewster3}.
In the literature the time independent response of detectors
with motion parallel to the boundary has been considered \cite{fsv:fordsvaitlyra,do:daviesott}.
We present the time dependent response of detectors with motion perpendicular to
the boundary.

The conical spacetime may be considered as the spacetime outside an
infinite, straight and zero radius cosmic string. The response of detectors
travelling parallel to such a string has been considered in \cite{ss:cosmic,ds:cosdet}.
We present the response of a detector approaching the string.

In section \ref{sec:dirac} we extend the model of Schlicht to the massless 
Dirac field both in Minkowski space and in the quotient spaces using the
automorphic field theory. For inertial and uniformly accelerated detectors
we recover the expected results for the transition
rate and power spectrum \cite{tk:takagi} on Minkowski space, and we discuss
these detectors on $M_0$.
Further we address the issue
as to whether or not such a detector can distinguish the 
two spin structures on $M_0$ and $M_-$ \cite{pl:langlois}.
Unfortunately on $M_-$ we find that our model is not sensitive to the
spin structure for any motions at constant $y$ and $z$.

Finally the responses on the boundary and conical spacetimes and the
higher dimensional generalizations of section \ref{sec:autodetector} are relevant also for the responses
of static detectors on the $\rp$ geon and inertial detectors in $\rp$ de Sitter 
space, via their associated global embedding Minkowski spaces (GEMS) \cite{dl:deser}.
Although until now the GEMS procedure has only been applied
to kinematical arguments we expect that at least in some 
cases the response of detectors in the original curved spaces and
the corresponding ones in their GEMS should be related in some way.
In section \ref{sec:Geon} we begin with a brief review of the GEMS literature. We 
then present an embedding of the Kruskal
manifold in a $7$-dimensional Minkowski space, introduced in \cite{gb:embed},
and a related embedding of the $\rp$ geon in a $7$-dimensional Minkowski space
with identifications. The Hawking temperature as seen by a static observer on the Kruskal manifold is
obtained by kinematical arguments from the related Unruh temperature of the observer in the 
embedding space. We then argue that on the $\rp$ geon the response of a static detector, in the
Hartle-Hawking-like vacuum, should be related to the response of the associated Unruh observer in the 
embedding space, in the Minkowski-like vacuum. This response is given by results in section \ref{sec:autodetector}.

In section \ref{sec-desitrp3} we consider inertial detectors in de Sitter and 
$\rp$ de Sitter space. We begin by introducing a causal detector and calculating the
response of a uniformly accelerated detector in de Sitter space, in the Euclidean vacuum, in a causal way. The
thermal result agrees with the literature \cite{dl:deser3}. Next we consider the
response of an inertial detector in $\rp$ de Sitter space, in the Euclidean-like vacuum with the motion
perpendicular to the distinguished foliation \cite{jk:desit}. The
response is seen to be identical to that of the uniformly accelerated detector on four-dimensional
Minkowski space with infinite plane boundary found in section \ref{sec:autodetector}.
On de Sitter and $\rp$ de Sitter space we also consider the response of
detectors via their $5$-dimensional GEMS. The thermal response in de Sitter space
has been given before \cite{dl:deser3}. The GEMS calculation on $\rp$ de Sitter space is
a particularly interesting one as we are able to present the 
calculation both in the curved space and in the embedding space.
It is found that the responses are qualitatively very similar.
This case should therefore be very useful in assessing the
validity of applying the GEMS procedure to cases involving
quotient spaces and time dependent detector responses.

We work throughout in natural units $\hbar=c=G=1$ and with metric signature 
$(+,-,\ldots,-)$. In $d$-dimensional Minkowski space, the spatial 
$(d-1)$-vectors are denoted by bold face 
characters $\mathbf{x}\in\R^{d-1}$ with $\cdot$ the usual scalar product 
in $\R^{d-1}$, while $d$-vectors 
(used occasionally) are given by an italic script $\mathit{x}$ 
with $\mathit{x}\cdot\mathit{y}=g_{\mu\nu}x^\mu{y^\nu}$, where 
$g_{\mu\nu}$ is the Minkowski metric.

\section{Causal detector in $d$ dimensions}
\label{sec-linmink}

In this section we introduce DeWitt's monopole detector \cite{d:dewitt}.
We discuss the use of spatial sampling functions
to regularize the correlation function 
and extend the regularization in \cite{sc:schlicht}, using a Lorentzian
sampling function, to $d$-dimensional Minkowski space $M$, $d>{2}$.
\footnote{The two-dimensional case may be dealt with in a similar way with the
added complication of the well known infrared divergence \cite{wh::whightman}. The 
correlation function contains an infinite constant term, which can be shown to not 
contribute to the response provided that the detector is switched on and off smoothly.
We shall not spell out the two-dimensional case further here.}

The model is that of a monopole detector, moving along a prescribed classical 
trajectory in $M$ and coupled to a massless scalar field $\phi$. The field has 
Hamiltonian $H_\phi$, and satisfies the massless Klein-Gordon equation. 
The free field operator is expanded in terms of a standard complete set of orthonormal solutions 
to the field equation as
\begin{equation}
\label{eqn:modeexpan}
\phi(t,\mathbf{x})=
\frac{1}{(2\pi)^{(d-1)/2}}
\int
\frac{d^{d-1}k}{(2\omega)^{1/2}}
\left(
a(\mathbf{k})e^{-i(\omega{t}-\mathbf{k}\cdot\mathbf{x})}
+a^{\dagger}(\mathbf{k})e^{i(\omega{t}-\mathbf{k}\cdot\mathbf{x})}
\right) \ ,
\end{equation}
where $(t,x_1,x_2,\ldots,x_{d-1})$ are usual Minkowski coordinates and, 
in the massless case, $\omega=|\mathbf{k}|$. The field is quantized by 
imposing, for the creation and annihilation operators, the usual commutation relations
\begin{equation}
\left[a(\mathbf{k}),a^{\dagger}(\mathbf{k})\right]
=
\delta^{d-1}(\mathbf{k}-\mathbf{k'}) \ . 
\end{equation}
The Minkowski vacuum $|0\rangle$ is the state annihilated by all 
the annihilation operators.

The detector is a quantum mechanical system with a set of energy
eigenstates $\{|0_D\rangle,|E_i\rangle\}$. It moves along 
a prescribed classical trajectory
$t=t(\tau)$, $\mathbf{x}=\mathbf{x}(\tau)$, where 
$\tau$ is the detector's proper time, and it couples to the scalar field via the 
interaction Hamiltonian
\begin{equation}
\label{eq:interhamil}
H_{\mathrm{int}}=
c
m(\tau)
\phi(\tau) \ ,
\end{equation}
where $c$ is a (small) coupling constant and $m(\tau)$ is the detector's monopole moment operator 
\cite{d:dewitt}. The evolution of $m(\tau)$ is given by
\begin{equation}
m(\tau)=
e^{iH_D\tau}m(0)e^{-iH_D\tau} \ . 
\end{equation}

Suppose that at time $\tau_0$ the detector and field are in the product state 
$|0,E_0\rangle=|0\rangle|E_0\rangle$, where $|E_0\rangle$ is a detector state with 
energy $E_0$. The probability that at time $\tau_1>\tau_0$ the detector is found in 
an excited state $|E_1\rangle$, regardless of the final state of the field,
is, to first order in perturbation theory
\begin{equation}
\sum_{\psi}
|\langle\psi,E_1|0,E_0\rangle|^2=
c^2
|\langle{E_1}|m(0)|E_0\rangle|^2
\int^{\tau_1}_{\tau_0}d\tau
\int^{{\tau}_1}_{{\tau}_0}d\tau'
e^{-i(E_1-E_0)(\tau-\tau')}\langle{0}|\phi(\tau)\phi(\tau')|0\rangle
 \ . 
\end{equation}
This expression has two parts. The sensitivity $c^2|\langle{E_1}|m(0)|E_0\rangle|^2$
depends only on the internal details of the detector and is not considered hereafter.
The ``response function''
\begin{equation}
F_{\tau_0,\tau_1}(\omega)=
\int^{\tau_1}_{\tau_0}d\tau
\int^{{\tau}_1}_{{\tau}_0}d\tau'
e^{-i\omega(\tau-\tau')}\langle{0}|\phi(\tau)\phi(\tau')|0\rangle
 \ ,
\end{equation}
where $\omega=E_1-E_0$ ($\omega>0$ for excitations and $\omega<0$ for de-excitations), 
does not depend on the internal details of the detector and so is
common for all such detectors. 

We now follow Schlicht \cite{sc:schlicht} and
change coordinates to $u=\tau$, $s=\tau-\tau'$ for $\tau'<\tau$
and $u=\tau'$, $s=\tau'-\tau$ for $\tau'>\tau$ and then 
differentiate with respect to $\tau_1$
to obtain an expression for the ``transition rate''
\begin{equation}
\label{eq:transition}
\dot{F}_{\tau_0,\tau}(\omega)=
2\int^{\tau-\tau_0}_{0}ds\,
Re
\left(
e^{-i\omega{s}}\langle{0}|\phi(\tau)\phi(\tau-s)|0\rangle
\right)
 \ ,
\end{equation}
where we have written $\tau_1=\tau$.
The transition rate is clearly causal in the sense that it does not depend on the state
of motion of the detector after time $\tau$ but only on times $\tau_0<\tau'<\tau$ which
label the past motion of the detector.

The correlation function $\langle{0}|\phi(x)\phi(x')|0\rangle$
in (\ref{eq:transition}) is the positive frequency Wightman function
which can be obtained from the expansion (\ref{eqn:modeexpan})
\begin{equation}
\label{eqn:integralwight}
\langle{0}|\phi(x)\phi(x')|0\rangle
=
\frac{1}{(2\pi)^{d-1}}
\int
\frac{d^{d-1}k}{2\omega}
e^{-i\omega(t-t')+i\mathbf{k}\cdot(\mathbf{x}-\mathbf{x}')}
\ .
\end{equation}
The integrals in (\ref{eqn:integralwight}) may be performed by first 
transforming to hyperspherical coordinates in $\mathbf{k}$-space.
The $|\mathbf{k}|$ integral requires regularization due to the
usual ultraviolet divergences found in quantum field theory.

The fundamental observation of reference \cite{sc:schlicht} is that if we regularize the divergences in
(\ref{eqn:integralwight}) using the usual $i\epsilon$-prescription, that is we introduce the
cut-off $e^{-\epsilon\omega}$, and then use it to compute the transition rate 
(\ref{eq:transition}) on a uniformly accelerated worldline with acceleration $1/\alpha$ and proper time $\tau$,
\begin{eqnarray}
t & = & \alpha\sinh(\tau/\alpha) \ ,
\nonumber
\\
\label{eq:accel}
x & = & \alpha\cosh(\tau/\alpha) \ ,
\end{eqnarray}
with the detector switched on in the infinite
past, $\tau_0=-\infty$, we obtain a time dependent and 
apparently unphysical result,
instead of the expected time independent thermal result 
(see e.g. \cite{d:dewitt,bd:book}).
Schlicht shows this in four dimensions by specific numerical and analytic 
calculations, and the general $d$ case follows similarly.

Schlicht's proposal is to consider an alternative regularization where the
monopole detector is considered as the limit of a detector with spatial 
extension which is rigid in the detector's rest frame. 
Instead of the interaction Hamiltonian (\ref{eq:interhamil})
with $\phi(\tau)=\phi(x(\tau))$ we consider (\ref{eq:interhamil}) with
the monopole moment operator coupled to the smeared field
\begin{equation}
\label{eq:smear}
\phi(\tau)=
\int
d^{d-1}\xi
\,
W_\epsilon(\mathbf{\xi})
\phi(x(\tau,\mathbf{\xi})) \ ,
\end{equation}
where $(\tau,\mathbf{\xi})$ are Fermi 
coordinates (see e.g.  \cite{MTW:grav}) and
$W_\epsilon(\mathbf{\xi})$ is a window sampling function
of characteristic length $\epsilon$.
In effect (\ref{eq:smear}) introduces a cut-off at short distances
of the order of $\epsilon$ in size.

We may choose a window function with infinite support, such
as that used in \cite{sc:schlicht}, if the window function
decreases sufficiently rapidly at large distances. Or
we may consider a detector of truly finite extent with the use
of a window function with compact support.
We require the window function to be normalised as 
\begin{equation}
\int
d^{d-1}\xi
\,
W_\epsilon(\mathbf{\xi})
=1 \ ,
\end{equation}
so that smearing a constant function
will 
return that constant value.
We also require the window function to approximate the
$(d-1)$-dimensional Dirac $\delta$ function, so that
in the limit as $\epsilon$ tends to $0$, (\ref{eq:smear})
formally gives the field value $\phi(x(\tau))$.
The window function hence staisfies $W_\epsilon(\mathbf{\xi})\approx{0}$
for $|\mathbf{\xi}|>>\epsilon$ and $W_\epsilon(\mathbf{\xi})\propto{\epsilon^{-(d-1)}}$
for $|\mathbf{\xi}|<<\epsilon$.

We now extend Schlicht's regularization of the Wightman function
(\ref{eqn:integralwight}) to $d$-dimensional
Minkowski space. Although a large number
of different window functions could be considered, the one chosen
in \cite{sc:schlicht} seems to be the easiest for obtaining a
closed expression for the Wightman function on an arbitrary
trajectory.

The $d$-dimensional analogue of the window function considered 
in \cite{sc:schlicht} is
\begin{equation}
\label{eq:shape}
W_{\epsilon}(\mathbf{\xi})
=
\frac{\Gamma[d/2]}{\pi^{d/2}}
\frac{\epsilon}{(\mathbf{\xi}^2+\epsilon^2)^{d/2}} \ .
\end{equation}
(\ref{eq:shape}) is sometimes referred to 
as a Lorentzian window or sampling function.
It approximates a $(d-1)$-dimensional Dirac $\delta$ function
and is suitably normalized.

Using (\ref{eq:smear}) and (\ref{eq:shape}),
we find
\begin{eqnarray}
\langle{0}|\phi(\tau)\phi(\tau')|0\rangle
& = &
\frac{1}{(2\pi)^{d-1}}
\int
\frac{d^{d-1}k}{2\omega}
\int
d^{d-1}\xi\,
W_{\epsilon}(\mathbf{\xi})
e^{-i({\omega}t(\tau,\mathbf{\xi})-\mathbf{k}\cdot\mathbf{x}(\tau,\mathbf{\xi}))}
\nonumber
\\
\label{eq:expan}
& & \times\int
d^{d-1}\xi'\,
W_{\epsilon}(\mathbf{\xi'})
e^{i({\omega}t(\tau',\mathbf{\xi'})-\mathbf{k}\cdot\mathbf{x}(\tau',\mathbf{\xi'}))} \ .
\end{eqnarray}
The integrals over $\mathbf{\xi}$ and $\mathbf{\xi'}$ in (\ref{eq:expan})
may be performed by transforming to hyperspherical coordinates and with
analogous arguments to those used in \cite{sc:schlicht} in four dimensions
we find
\begin{equation}
\label{eq:correlation}
\langle{0}|\phi(\tau){\phi}(\tau')|0\rangle=
\frac{1}{(2\pi)^{d-1}}
\int
\frac{d^{d-1}k}{2\omega}
e^{-i\omega(t-t'-i\epsilon(\dot{t}+\dot{t}'))
+i\mathbf{k}\cdot(\mathbf{x}-\mathbf{x'}-i\epsilon(\dot{\mathbf{x}}+\dot{\mathbf{x}}'))} \ .
\end{equation}

The regularization introduced by the smearing of the field in the detector's rest frame
may be viewed as an ultraviolet cut-off where high frequencies as seen by the detector
are cut-off (as opposed to the high frequencies as seen by an inertial observer
which are cut-off by the usual $i\epsilon$-prescription).
The frequency is the time component of the $4$-momentum. Given that the $4$-momentum
in the usual Minkowski frame is $(\omega,\mathbf{k})^{\top}$ we find via a straightforward
calculation that the time component of this $4$-vector in the Fermi frame at time
$\tau$ is $\omega\dot{t}(\tau)-\mathbf{k}\cdot\dot{\mathbf{x}}(\tau)$. The regularization
in (\ref{eq:correlation}) is seen to be equivalent to an exponential cut-off of the high 
frequency modes, as seen in the detector's frame, at times $\tau$ and $\tau'$.

The integrals in (\ref{eq:correlation}) may also be done by transforming to hyperspherical 
coordinates. 
The result is
\begin{eqnarray}
& \langle{0}|\phi(\tau)\phi(\tau')|0\rangle=
\frac{\Gamma[d/2-1]}{4\pi^{d/2}}
\frac{1}{A^{d/2-1}} \ , \nonumber
\\
& A=\left[i^2(t(\tau)-t(\tau')-i\epsilon(\dot{t}(\tau)+\dot{t}(\tau')))^2
+(\mathbf{x}(\tau)-\mathbf{x}(\tau')-i\epsilon(\dot{\mathbf{x}}(\tau)+\dot{\mathbf{x}}(\tau')))^2\right] \ .
\nonumber
\\
\label{eq:corre}
\end{eqnarray}

We note here that following a similar calculation to that given above
the usual $i\epsilon$ regularization of the correlation function
may also be obtained by considering a spatially smeared field. The difference from the above
regularization is that the smearing is done in the Minkowski 
reference frame, that is, always with respect to inertial observers, and
not in the Fermi frame. This model detector is thus not rigid in its 
rest frame.

First we consider the transition rate (\ref{eq:transition}) with the correlation 
function (\ref{eq:corre})
for a detector following an inertial trajectory. Consider therefore the 
trajectory $t=\tau$, $\mathbf{x}=\mathrm{constant}$,  
where $-\infty<\tau<\infty$.
If the detector is switched on in the infinite past, $\tau_0=-\infty$,
the transition rate is 
\begin{equation}
\dot{F}_\tau(\omega)=
\frac{\Gamma[d/2-1]}{4\pi^{d/2}}
\int^{\infty}_{-\infty}ds\,
\frac{e^{-i\omega{s}}}
{[i^2(s-2i\epsilon)^2]^{d/2-1}} \ .
\end{equation}
Note that we obtain the same expression in this case if we use instead the
usual correlation function i.e. with
the $i\epsilon$ regularization. The integral may be done by residues 
and the result as $\epsilon\rightarrow{0}$ is 
\begin{equation}
\label{eqn:inertrespons}
\dot{F}_\tau(\omega)=
\frac{\Gamma[d/2-1](-\omega)^{d-3}}{2\pi^{d/2-1}(d-3)!}
\Theta(-\omega) \ ,
\end{equation}
where $\Theta(-\omega)$ is the Heaviside step function. 
As expected the transition rate vanishes for $\omega>0$, indicating that an inertial detector
is not excited by the Minkowski vacuum. For $\omega<0$ the transition rate is non-zero due to
the possibility of spontaneous emission by the detector.

Next consider the transition rate for a spatially extended detector
whose centre follows the uniformly accelerated worldline (\ref{eq:accel}).
The transition rate is 
\begin{equation}
\label{transaccel}
\dot{F}_\tau(\omega)=
\frac{\Gamma[d/2-1]}{i^{d-2}(4\pi)^{d/2}}
\int^{\Delta\tau}_{-\Delta\tau}
ds\,
\frac{e^{-i\omega{s}}}{\left(
\alpha\sinh\left(\frac{s}{2\alpha}\right)-i\epsilon\cosh\left(\frac{s}{2\alpha}\right)
\right)^{d-2}} \ .
\end{equation}
In the case of a detector switched on in the infinite past, $\tau_0=-\infty$,
the integral here may be done by residues.
The transition rate is independent of $\tau$ along the trajectory
and is given by
\begin{eqnarray}
\dot{F}_\tau(\omega)= 
&  &
\frac{\pi}{2^{d-2}\pi^{(d-1)/2}\alpha^{d-3}\Gamma((d-1)/2)}
\nonumber
\\
\label{thermal}
& \times &
\left\{
\begin{array}{ll}
\frac{\alpha\omega}{\left(e^{2\pi\omega\alpha}-1\right)}
\prod_{k=1}^{(d-4)/2}
\left(
\left(\frac{d-2}{2}-k\right)^2
+\alpha^2\omega^2
\right)
&
d\;\; \textrm{even}
\\
\frac{1}{\left(e^{2\pi\omega\alpha}+1\right)}
\prod_{k=1}^{(d-3)/2}
\left(
\left(\frac{d-2}{2}-k\right)^2
+\alpha^2\omega^2
\right)
&
d\;\; \textrm{odd}
\end{array}
\right. \ ,
\end{eqnarray}
where for $d=3$ and $d=4$ the products $\prod_{k=1}^{(d-3)/2}\left(
\left(\frac{d-2}{2}-k\right)^2
+\alpha^2\omega^2
\right)$ and \\
$\prod_{k=1}^{(d-4)/2}\left(
\left(\frac{d-2}{2}-k\right)^2
+\alpha^2\omega^2
\right)$ in (\ref{thermal}) are both $1$.
The transition rate (\ref{thermal}) is 
as in the literature \cite{sr:sriramk}.
It is thermal with characteristic temperature $T=1/(2\pi\alpha)$
in the sense that is satisfies the KMS condition
\begin{equation}
\label{eqn:kmscondition}
\dot{F}_\tau(\omega)=
e^{-\omega/T}
\dot{F}_\tau(-\omega) \ ,
\end{equation} 
at that temperature (see e.g \cite{tk:takagi}).
Further (\ref{thermal}) contains the expected ``apparent'' statistics inversion 
as we go from odd to even dimensions.

\section{Massive scalar field}
\label{sec:massive}

In this section we compute the correlation function for the smeared field operator (\ref{eq:smear}) for
a massive scalar field. 

The detector model is as in section \ref{sec-linmink}.
The field is expanded in modes as in (\ref{eqn:modeexpan}) but now with $\omega=\left(\mathbf{k}^2+m^2\right)^{1/2}$,
where $m$ is the field mass. The transition rate for the detector is given by (\ref{eq:transition}),
where 
$\phi(\tau)$ is the smeared field operator (\ref{eq:smear}). The correlation 
function is given by the expression (\ref{eq:expan}) and following
an identical calculation to that which leads to (\ref{eq:correlation}) we find
\begin{equation}
\label{eq:correlationmass}
\langle{0}|\phi(\tau){\phi}(\tau')|0\rangle=
\frac{1}{(2\pi)^{d-1}}
\int
\frac{d^{d-1}k}{2\omega}\,
e^{-i\omega(t-t'-i\epsilon(\dot{t}+\dot{t}'))
+i\mathbf{k}\cdot(\mathbf{x}-\mathbf{x'}-i\epsilon(\dot{\mathbf{x}}+\dot{\mathbf{x}}'))} \ ,
\end{equation}
with $\omega=\left(\mathbf{k}^2+m^2\right)^{1/2}$.

The integrals here
may again be done by moving to hyperspherical coordinates in $k$-space.
We restrict ourselves now to the case $d=4$ (although the arbitrary $d$ case
follows similarly). After performing the
angular integrals, we find
\begin{equation}
\langle{0}|\phi(\tau)\phi(\tau')|0\rangle
=  
\frac{1}{(2\pi)^2R}
\int_0^\infty
dk\,
\frac{k}{\left(k^2+m^2\right)^{1/2}}
\sin(kR)
e^{-i\left(k^2+m^2\right)^{1/2}{(t-t'-i\epsilon(\dot{t}+\dot{t}'))}} \ ,
\end{equation}
where $R=\sqrt{(\mathbf{x}-\mathbf{x'}-i\epsilon(\dot{\mathbf{x}}+\dot{\mathbf{x}}'))^2}$. This may be written
as
\begin{equation}
\label{eqn:massinteq}
\langle{0}|\phi(\tau)\phi(\tau')|0\rangle
=  
\frac{-1}{8\pi^2R}
\partial_R
\int_{-\infty}^\infty
dk\,
\frac{1}{\left(k^2+m^2\right)^{1/2}}
e^{-i(\left(k^2+m^2\right)^{1/2}{(t-t'-i\epsilon(\dot{t}+\dot{t}'))}-kR)}
 \ .
\end{equation}
We now change variables by $k=m\sinh{\theta}$, so that $\omega=m\cosh{\theta}$ and
\begin{equation}
\langle{0}|\phi(\tau)\phi(\tau')|0\rangle
=  
\frac{-1}{8\pi^2R}
\partial_R
\int_{-\infty}^\infty
d\theta\,
e^{im(R\sinh{\theta}-(t-t'-i\epsilon(\dot{t}+\dot{t}'))\cosh{\theta})}
 \ .
\end{equation}

For the detector trajectory we consider only timelike worldlines,
$(t-t')>(\mathbf{x}-\mathbf{x}')$. We distinguish two cases.
Firstly, for $(t-t')>0$, we make the substitution $(t-t'-i\epsilon(\dot{t}+\dot{t}'))=
\sqrt{\lambda}\cosh{\theta_0}$, $R=\sqrt{\lambda}\sinh{\theta_0}$, with
$\lambda=(t-t'-i\epsilon(\dot{t}+\dot{t}'))^2-R^2$. We find
\begin{equation}
\langle{0}|\phi(\tau)\phi(\tau')|0\rangle
=  
\frac{-1}{8\pi^2R}
\partial_R
\int_{-\infty}^\infty
d\theta\,
e^{-im\sqrt{\lambda}\cosh(\theta_0-\theta)}
 \ .
\end{equation} 
Now we note
\begin{equation}
\label{eqn:K0}
K_0(z)=\frac{1}{2}\int_{-\infty}^{\infty}dt\,
e^{-z\cosh{t}} \ ,
\end{equation}
valid for $Re(z)>0$ \cite{gr:inttables}, where $K_0$ is a modified Bessel 
function, and we may show for a timelike worldline that $Im(\sqrt{\lambda})<0$.
Hence
\begin{eqnarray}
\langle{0}|\phi(\tau)\phi(\tau')|0\rangle
& = &   
\frac{-1}{4\pi^2R}
\;\partial_R\left[
K_0\left(im\sqrt{\lambda}\right)\right]
\nonumber
\\
\label{eqn:masscorr}
& = &
-\frac{im}{4\pi^2\sqrt{\lambda}}
\;K_1\left(im\sqrt{\lambda}\right) \ ,
\end{eqnarray} 
as $\partial_z{K_0(z)}=K_1(z)$.
The case $(t-t')<0$ is similar except we make the
change of variables $(t-t'-i\epsilon(\dot{t}+\dot{t}'))=
-\sqrt{\lambda}\cosh{\theta_0}$, $R=\sqrt{\lambda}\sinh{\theta_0}$
again we may use the integral representation of $K_0$ and we obtain the
same result (\ref{eqn:masscorr}).

The massless limit is easily checked. We have near $z=0$, $K_1(z)=1/z$
\cite{gr:inttables} and so the correlation function (\ref{eqn:masscorr})
agrees with that of Schlicht \cite{sc:schlicht} in this limit.
Had we used the field operator without smearing we would have obtained (\ref{eqn:massinteq}) with
$\epsilon=0$. The usual regularization procedure, as with the massless
case, would then be to introduce a cut-off in the high frequency modes by $t\rightarrow{t-i\eta}$, where
$\eta$ is small. The result would be (\ref{eqn:masscorr}) but with
$\lambda=(t-t'-i\eta)^2-|\mathbf{x}-\mathbf{x}'|^2$.


\section{Detector on Quotient spaces}
\label{sec:autodetector}

In this section we adapt the detector model of section \ref{sec-linmink} to 
spacetimes built as quotients $M/\Gamma$ of Minkowski
space under certain discrete symmetry groups $\Gamma$.
In particular we calculate responses
on $M_0$, $M_-$ \cite{lm:geon,pl:langlois,pl:langlois2}, Minkowski space with an infinite plane boundary 
and certain conical spacetimes.

As these
quotient spaces do not have infinite spatial sections in all directions,
it does not directly make sense to consider 
a detector with infinite spatial extent as used in (\ref{eq:smear}).
We shall argue however that we may introduce a detector similar to that of 
section \ref{sec-linmink} by working with automorphic fields on $M$~\cite{bD:auto,bd:em}.

Consider Minkowski space $M$ in $d$ dimensions, and consider the quotient 
space $M/\Gamma$ where $\Gamma$ is some discrete isometry group. $|\Gamma|$
may be infinite (as indeed is the case on $M_0$ and $M_-$) which will
mean that some of the following expressions remain formal in those cases.
We will find however that these formalities do not interfere as in any calculations
done the infinities and the formally vanishing normalization factors will cancel to
give finite results.

The automorphic 
field $\hat{\phi}$ is constructed from the ordinary field ${\phi}$ as the sum
\begin{equation}
\label{eqn:autofield}
\hat{\phi}(x)
:=
\frac{1}{\left(\sum_{\gamma\in\Gamma}p(\gamma)^2\right)^{1/2}}
\sum_{\gamma\in\Gamma}p(\gamma)\phi(\gamma^{-1}x) \ ,
\end{equation}
where $p(\gamma)$ is a representation of $\Gamma$ in $SL(\R)\simeq\{1,-1\}$.
The normalization in (\ref{eqn:autofield}) has been chosen so that, at equal times
\begin{equation}
\left[\hat{\phi}(x),\dot{\hat{\phi}}(x')\right]
=
i\delta^{(d-1)}(x-x')
+\textrm{image terms} \ .
\end{equation}
The two point function for the automorphic field is then given by the 
method of images as
\begin{equation}
\label{eqn:automorcorre}
\langle{0}|\hat{\phi}(x)\hat{\phi}(x')|0\rangle
=
\sum_{\gamma\in\Gamma}p(\gamma)\langle{0}|\phi(x)\phi(\gamma^{-1}x')|0\rangle \ ,
\end{equation}
where $\langle{0}|\phi(x)\phi(x')|0\rangle$ is the usual correlation 
function on Minkowski space.

As a model of a particle detector on $M/\Gamma$, we introduce on $M$ a
detector linearly coupled to the automorphic field by
\begin{equation}
\label{eqn:hamiauto}
H_{\mathrm{int}}=
c
m(\tau)
\hat{\phi}(\tau) \ ,
\end{equation}
with\footnote{If we considered here $\hat{\phi}(\tau)
=\hat{\phi}(x(\tau))$ with the usual $i\epsilon$ regularization,
again we would find as Schlicht does in Minkowski space an unphysical result
for the response of the uniformly accelerated detector on these spacetimes. 
This is most easily seen by considering that
the $\gamma=I$ term (where $I$ is the identity element) in (\ref{eqn:automorcorre}) 
is that found in Minkowski space.}
\begin{equation}
\label{eqn:autointham}
\hat{\phi}(\tau)
=
\int
d^{d-1}\xi
\,
W_{\epsilon}(\mathbf{\xi})
\hat{\phi}(x(\tau,\mathbf{\xi})) \ .
\end{equation}
One might ask why we have not included in (\ref{eqn:hamiauto}) image terms under
$\Gamma$ (that is one term for each image of the detector). There are two obvious ways
in which such terms could be included. Firstly we could consider each image
term in the sum to be weighted by the representation $p(\gamma)$. In certain situations, such as
on Minkowski space with an infinite plane boundary with Dirichlet boundary conditions
this would however lead to the detector and its image terms cancelling each other
to give a vanishing interaction Hamiltonian. 
Alternatively we could consider image terms without the representation weights.
Then each image term would be equal to that in (\ref{eqn:hamiauto}) and so we 
would obtain the same results with an overall (possibly infinite) normalisation factor, which can 
be absorbed in the coefficient $c$. We therefore work with (\ref{eqn:hamiauto}).

From (\ref{eqn:hamiauto}), a discussion analogous to that in section \ref{sec-linmink}
leads to the transition rate 
(to first order in perturbation theory)
\begin{equation}
\label{eqn:autotrans}
\dot{F}_{\tau}(\omega)=
2\int^{\infty}_{0}ds
\,
Re
\left(
e^{-i\omega{s}}\langle{0}|\hat{\phi}(\tau)\hat{\phi}(\tau-s)|0\rangle
\right) \ ,
\end{equation}
where the correlation function for
the automorphic field in (\ref{eqn:autotrans}), $\langle{0}|\hat{\phi}(\tau)\hat{\phi}(\tau-s)|0\rangle$,
is given by the method of images applied to (\ref{eq:corre}).

\subsection{$M_0$}

$M_0$ is a quotient of $4$-dimensional Minkowski space where the quotienting group
$\Gamma$ is that generated by the isometry $J_0:(t,x,y,z)\mapsto(t,x,y,z+2a)$.
The transition rate for our detector is given by
(\ref{eqn:autotrans}) with 
\begin{equation}
\label{eqn:corronm0}
\langle{0}|\hat{\phi}(\tau)\hat{\phi}(\tau')|0\rangle=\sum_{n\in{\Z}}\eta^n\langle{0}|\phi(\tau)\phi(J_0^n\tau')|0\rangle
 \ ,
\end{equation}
where $\eta=+1,(-1)$ are the representations of $\Gamma$, labelling untwisted (twisted)
fields respectively. 

\subsubsection{Inertial detector on $M_0$}

Consider first a detector following the inertial trajectory
\begin{eqnarray}
t=\tau(1-v^2)^{-1/2} \ , & z=\tau{v}(1-v^2)^{-1/2}  \ ,\nonumber \\
x=x_0  \ , & y=y_0 \ ,  
\end{eqnarray}
where velocity $-1<v<0$, $-\infty<\tau<\infty$ is the detector's proper time and 
$x_0$, $y_0$ are constants.
Substituting the trajectory into (\ref{eqn:corronm0}) and 
then (\ref{eqn:autotrans}), we find that
the transition rate for a detector switched on in the 
infinite past 
reads
\begin{equation}
\dot{F}_\tau(\omega)=
-\frac{1}{4\pi^2}
\sum_{n=-\infty}^{\infty}
\eta^n
\int^\infty_{-\infty}
ds
\,
\frac{e^{-i\omega{s}}}{(s-2i\epsilon)^2+4nav(s-2i\epsilon)(1-v^2)^{-1/2}-(2na)^2}
\ .
\end{equation}
The integral may be done by residues. The result is
\begin{equation}
\dot{F}_\tau(\omega)=
-\frac{(1-v^2)^{1/2}}{4\pi{a}}
\sum_{n=-\infty}^{\infty}
\eta^n
\frac{\sin(2\omega{n}a(1-v^2)^{-1/2})}{n}
e^{\frac{2\omega{n}{a}vi}{(1-v^2)^{1/2}}}\Theta(-\omega) \ .
\end{equation}
As on Minkowski space the transition rate vanishes for 
$\omega>0$ while it is non-zero for $\omega<0$ although the
rate of spontaneous emission is altered from the Minkowski rate
due to the non-trivial topology.
The transition rate depends on velocity $v$ due to the absence of
a boost Killing vector in the $z$ direction on $M_0$.

Consider now the limit as $v\rightarrow{0}$. By the
isometries of $M_0$ this gives the response of an inertial 
detector that may have arbitrary velocity in the $x$ or $y$ directions.
Then
\begin{equation}
\label{eqn:transm0scalarinert}
\dot{F}_\tau(\omega)=
\left(
-\frac{\omega}{2\pi}
+\frac
{1}
{2\pi{a}}
\sum_{n=1}^{\infty}
\eta^n
\frac{1}
{n}
\sin(-2n\omega{a})
\right)
\Theta(-\omega) \ .
\end{equation}
In the case 
of an untwisted field, $\eta=1$, the summation in (\ref{eqn:transm0scalarinert}) is recognized as 
the Fourier series of the $2\pi$-periodic
function that on the
interval $(0,2\pi)$ takes the form $f(-2\omega{a})=\frac{1}{2}(\pi+2\omega{a})$ (see e.g. \cite{zw:zwillinger}).
We hence find
\begin{equation}
\label{eqn:inerm0}
\dot{F}_\tau(\omega)=
\frac
{\left(\left[\frac{-\omega{a}}{\pi}\right]+\frac{1}{2}\right)}
{2a}
\Theta(-\omega) \ ,
\end{equation}
where $[x]$ denotes the integer part of $x$. 

For a twisted field, $\eta=-1$, we note
$(-1)^n\sin(nx)=\sin(n(x+\pi))$ and find
\begin{equation}
\label{eqn:inerm2}
\dot{F}_\tau(\omega)=
\frac
{\left[\frac{-\omega{a}}{\pi}+\frac{1}{2}\right]}
{2a}
\Theta(-\omega) \ .
\end{equation}

\subsubsection{Uniformly accelerated detector}

If we consider a detector following the worldline of uniform 
acceleration (\ref{eq:accel}) we obtain, again for the detector
switched on at $\tau_0=-\infty$,
\begin{equation}
\label{M0response}
\dot{F}_\tau(\omega)=\frac{\omega}{2\pi(e^{2\pi\alpha\omega}-1)}
\left(
1+
\sum_{n=1}^{\infty}
\eta^n
\frac{\sin{[2\alpha\omega{\mathrm{arc}}\sinh{(na/\alpha)}]}}{na\omega\sqrt{1+n^2a^2/\alpha^2}}
\right) \ ,
\end{equation}
where the integral in the transition rate has again been done by residues.
The result agrees with that obtained by Louko and Marolf in \cite{lm:geon}.
The response is independent of $\tau$. Moreover it is thermal in the
sense that it satisfies the KMS condition (\ref{eqn:kmscondition}).
There is however a break
from the purely Planckian form found on Minkowski space.

\subsection{Minkowski space with an infinite plane boundary}
\label{sec:minkbound}

In this subsection we consider a detector on $d$-dimensional Minkowski space, $d>2$, 
with Minkowski coordinates $(t,x_1,\ldots,x_{d-1})$,
with an infinite boundary at $x_1=0$. 
Detectors on this spacetime (in particular in the case 
of $4$-dimensions)
have been considered by a number of authors (see e.g. \cite{dlo:detecbound,sz:suzuki})\footnote{The case of 
Dirichlet boundary conditions for $d=4$ has been the focus
of some study recently on the response of detectors to negative energy~\cite{do:daviesott}.
The authors of \cite{do:daviesott} considered the response of a finite time detector travelling inertially
parallel to the boundary. They found that the negative energy outside the boundary
has the effect of decreasing the excitations which are present even in Minkowski space 
due to the switching of the detector.}.
However there has not been any presentations (as far as the author is aware)
of the time dependent response for an inertial or uniformly accelerated observer
who approaches the boundary from infinity.
The results in this section are interesting also as a preliminary calculation 
before the response of detectors on $M_-$ is considered. Many of the features
seen here will also be observed there, but in a simpler context as there is no compact
direction and therefore the image sum is finite.
Further we will see in section \ref{sec:rp3} that the response of a uniformly accelerated
detector on four-dimensional Minkwoski space with an infinite plane boundary 
is very closely related to the response of an 
inertial detector coupled to a conformal scalar field
in $\rp$ de Sitter space~\cite{jk:desit}.

Again the discussion of 
section \ref{sec:autodetector} follows through and the 
transition rate is given by (\ref{eqn:autotrans})
where now the automorphic correlation function is
\begin{equation}
\label{eqn:correonbound}
\langle{0}|\hat{\phi}(\tau)\hat{\phi}(\tau')|0\rangle
=
\sum_{n=0,1}\beta^n\langle{0}|\phi(\tau)\phi(J_b^n\tau')|0\rangle \ ,
\end{equation}
where
$J_b:(t,x_1,x_2,\ldots,x_{d-1})\mapsto(t,-x_1,x_2,\ldots,x_{d-1})$ and
$\beta=+1,(-1)$, which label Neumann and (Dirichlet) boundary conditions
respectively.
We note here that on four-dimensional Minkowski space with boundary at $x=0$ 
the renormalized expectation values $\langle{0}|T_{\mu\nu}|0\rangle$
of the energy-momentum tensor for the minimally coupled massless 
scalar field in the vacuum state induced by the Minkowski vacuum
are \cite{bd:book,pl:langlois2}
\begin{equation}
\label{eqn:tmunu2}
\langle0|T_{tt}|0\rangle
=
-\langle0|T_{yy}|0\rangle
=
-\langle0|T_{zz}|0\rangle
=\beta\frac{1}{16{\pi^2}x^4}
 \ ,\;\;
\langle0|T_{xx}|0\rangle
=0 \ .
\end{equation}

\subsubsection{Inertial detector}

\begin{figure}[htbp]
\includegraphics[angle=0, width={4in}]{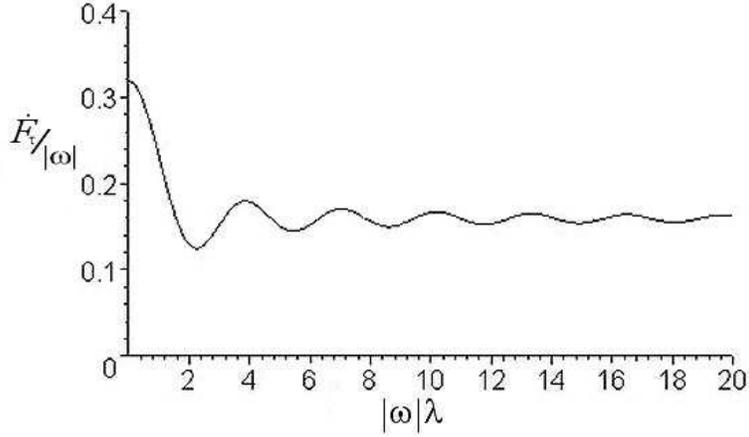}
\centering
\caption[Transition rate of inertial detector on $M$ with boundary, motion parallel to boundary, $\omega<0$.]{Transition rate for inertial detector moving parallel to the boundary, as
a function of distance $\lambda$ from the boundary. We have taken $\omega<0$ and plotted
$\dot{F}_{\tau}(\omega)/|\omega|$ against $|\omega|\lambda$ for
Neumann boundary conditions.}
\label{fig:inertialparal}
\end{figure}

Firstly we consider an inertial detector with motion
parallel to the boundary. Due to the isometries of the
spacetime we may consider, without loss of generality, the trajectory
$t=\tau$, $x_1=\lambda$, $x_i=0$ for $1<i\leq{d-1}$
where $\lambda$, the distance from the boundary, is constant.
The transition rate contains two terms. The first one comes 
from the first term in the boundary space correlation
function (\ref{eqn:correonbound}) and equals the corresponding 
transition rate on Minkowski space. For a detector
switched on in the infinite past this is given by (\ref{eqn:inertrespons}).
The second term, comes from the image term in (\ref{eqn:correonbound}),
and is given by
\begin{equation}
\dot{F}_{B\tau}(\omega)=
\frac{\beta\Gamma(d/2-1)}
{4\pi^{d/2}}
\int_{-\infty}^{\infty}
ds\,
\frac{e^{-i\omega{s}}}
{[i^2(s-2i\epsilon)^2+(2\lambda)^2]^{d/2-1}} \ .
\label{eq:boundparttrans}
\end{equation}
The integral in (\ref{eq:boundparttrans}) can be evaluated by residues.
Specialising to $4$ dimensions, the total transition rate including the Minkowski 
part is found to be
\begin{equation}
\label{eqn:inertialparallel}
\dot{F}_{\tau}(\omega)=
\left(-\frac{\omega}{2\pi}
-\frac{\beta}
{4\pi\lambda}
\sin(2\omega\lambda)
\right)
\Theta(-\omega) \ .
\end{equation}
In figure \ref{fig:inertialparal} we plot $\dot{F}_{\tau}(\omega)/|\omega|$ against
$|\omega|\lambda$ for Neumann boundary conditions.
On the boundary the rate
is twice that in Minkowski space, while far from the boundary the rate becomes that
on Minkowski space.
For Dirichlet boundary conditions the transition rate vanishes on the
boundary as expected.
Our results agree with those in~\cite{dlo:detecbound}.

Next, restricting ourselves to $d=4$ again, we consider an 
inertial detector approaching the boundary from 
infinity following the worldline 
\begin{eqnarray}
\label{eqn:inertraj}
t=\tau(1-v^2)^{-1/2} \ , & x=\tau{v}(1-v^2)^{-1/2}  \ ,\nonumber \\
y=y_0  \ , & z=z_0 \ ,  
\end{eqnarray}
with $-1<v<0$ and $-\infty<\tau<0$.
We expect the response in this case
to be dependent on proper time $\tau$, as the boundary breaks the
translation invariance of Minkowski space in the $x$-direction.
 
Again the transition rate is in two parts. The 
Minkowski part (the $n=0$ term) will lead again to the Minkowski space rate
(\ref{eqn:inertrespons}). For a detector switched on the the infinite past
this part reads
\begin{equation}
\dot{F}_{M\tau}(\omega)=
-\frac{\omega}
{2\pi}
\Theta(-\omega) \ ,
\end{equation}
where $\Theta(-\omega)$ is a step function, and $M$ denotes
that this is the Minkowski term. The Minkowski term of course 
is independent of $\tau$.
It is the image term in the correlation function which leads to a $\tau$-dependent 
result. The image part of the transition rate is
\begin{equation}
\label{eqn:boundpartmboundinet}
\dot{F}_{B\tau}(\omega)=
-\frac{\beta}{2\pi^2}
(1-v^2)
\lim_{\epsilon\rightarrow{0}}
\int^{\infty}_{0}ds\,
Re
\left(
\frac
{e^{-i\omega{s}}}
{(s-2i\epsilon)^2-v^2({s-2\tau})^2}
\right) \ .
\end{equation}
The integral may be evaluated, using some contour arguments, in terms
of sine and cosine integrals. We find
\begin{eqnarray}
\dot{F}_\tau(\omega) & = & -\frac{\beta}{2\pi^2(b+c)}
(
-\mathrm{Ci}(b|\omega|)\cos(b|\omega|)
-\mathrm{si}(b|\omega|)\sin(b|\omega|)
\nonumber
\\
\label{eqn:inertonbound}
&  & +\mathrm{Ci}(c|\omega|)\cos(c|\omega|)
+\mathrm{si}(c|\omega|)\sin(c|\omega|)
+2\pi\sin(b\omega)\Theta(-\omega)
) \ ,
\end{eqnarray}
where $\mathrm{Ci}$, $\mathrm{si}$ are the cosine and shifted sine integrals 
\cite{as:abromowitz}, $b=2v\tau/(1+v)$ and $c=2v\tau/(1-v)$.

\begin{figure}[htbp]
\includegraphics[angle=0, width={4in}]{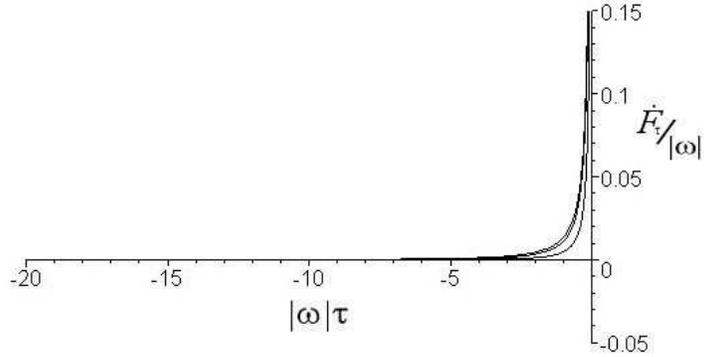}
\centering
\caption[Transition rate of inertial detector on $M$ with boundary, motion perpendicular to boundary, $\omega>0$.]{Transition rate for inertial detector approaching boundary with Neumann boundary conditions
and $\omega>0$.
$\dot{F}_\tau(\omega)/|\omega|$ is plotted against $|\omega|\tau$ 
for $v=-1/2$ (lower curve near the axis), $v=-1/3$ and $v=-1/4$ (upper curve).}
\label{fig:inertialwg0}
\end{figure}

In figure \ref{fig:inertialwg0} we plot 
$\dot{F}_\tau(\omega)/|\omega|$ against $|\omega|\tau$ for $\omega>0$, Neumann boundary conditions 
and for various values of $v$.
We note that for $\omega>0$ the transition rate is non-zero, in contrast 
to the response of an inertial detector travelling parallel to the boundary,
and diverges as the boundary is reached.

\begin{figure}[htbp]
\includegraphics[angle=0, width={4in}]{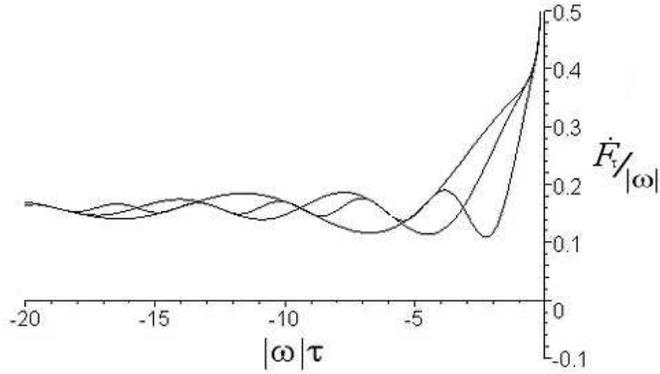}
\centering
\caption[Transition rate of inertial detector on $M$ with boundary, motion perpendicular to boundary, $\omega<0$.]{Transition rate for inertial detector approaching boundary with Neumann boundary conditions
and $\omega<0$.
$\dot{F}_\tau(\omega)/|\omega|$ is plotted against $|\omega|\tau$ for
$v=-1/2$ (lower curve), $v=-1/3$ and $v=-1/4$ (upper curve).}
\label{fig:inertialwl0}
\end{figure}

For $\omega<0$ recall that the Minkowski part of the transition rate is non zero.
We plot the total rate for $\omega<0$ in figure \ref{fig:inertialwl0}.
In both cases the response depends on the velocity, as expected since
there is no boost isometry in the $x$-direction.
Further in both cases we may show that the divergence at $x=0$ goes as 
$1/\tau$ and so is weaker than that in the energy expectation values (\ref{eqn:tmunu2}).
It can also be verified that the transition rate dies off at $\tau=-\infty$ as $O(1/\tau^3)$.
Further numerical evidence suggests that the divergences persist
for a detector that is switched on at a finite time.

\subsubsection{Uniformly accelerated detectors}
\label{sec:autoacelmbound}

Consider now a uniformly accelerated detector with acceleration parallel to the boundary and
switched on in the infinite past. We may consider
without loss of generality the worldline
\begin{eqnarray}
t & = & \alpha\sinh(\tau/\alpha) \ ,
\nonumber
\\
x_1 & = & \lambda \ ,
\nonumber
\\
x_2 & = & \alpha\cosh(\tau/\alpha) \ ,
\nonumber
\\
x_i & = & 0 \ , \;\;\;\;\;\;\;\;\;\;\;\;\;\;\;\;2<i\leq{d-1} \ .
\label{eqn:accelboundparr}
\end{eqnarray}

The response again is in two parts.
The first term in (\ref{eqn:correonbound}) leads to the
thermal transition rate an accelerated detector on Minkowski space
(\ref{thermal}). The image part of the correlation function
on the worldline (\ref{eqn:accelboundparr}) is
\begin{equation}
\label{cronaccelbound}
\langle{0}|\phi(\tau)\phi(J_B\tau')|0\rangle=
\frac{\beta\Gamma[d/2-1]}{i^{d-2}4\pi^{d/2}}
\frac{1}{\left(
4\left(\alpha\sinh(\frac{\tau-\tau'}{2\alpha})-i\epsilon\cosh(\frac{\tau-\tau'}{2\alpha})\right)^2
-(2\lambda)^2
\right)^{d/2-1}} \ ,
\end{equation}
which as expected is invariant under $\tau$ translations. Restricting now
to $4$ dimensions, the boundary part of the transition rate is
\begin{equation}
\label{eqn:boundaryaccelparra}
\dot{F}_{B\tau}(\omega)=
-\frac{\beta}
{4\pi^2}
\int^{\infty}_{-\infty}ds\,
\frac{e^{-i\omega{s}}}
{\left(
4\left(\alpha\sinh(\frac{s}{2\alpha})-i\epsilon\cosh(\frac{s}{2\alpha})\right)^2
-(2\lambda)^2
\right)} \ .
\end{equation}
The integral can be done by residues. 
The result is
\begin{equation}
\label{eqn:boundaryaccelparra2}
\dot{F}_{B\tau}(\omega)=
\frac{\beta}
{4\pi}
\frac{\alpha}{\lambda(\alpha^2+\lambda^2)^{1/2}}
\frac{1}{(e^{2\pi\omega\alpha}-1)}
\sin(2\omega\alpha\mathrm{arcsinh}(\lambda/\alpha)) \ ,
\end{equation}
which agrees with~\cite{dlo:detecbound}.
The response is thermal in the sense that it satisfies the KMS
condition at temperature $T=(2\pi\alpha)^{-1}$.

Now let us consider the uniformly accelerated worldline (\ref{eq:accel}).
The acceleration is now perpendicular to the boundary.
We begin by considering the detector switched on in the
infinite past. 
The Minkowski part of the correlation function again leads
to the thermal response (\ref{thermal}). The image term
on worldline (\ref{eq:accel}) gives
\begin{equation}
\label{eqn:imtermperpacc}
\langle{0}|\phi(\tau)\phi(J_B\tau')|0\rangle
=
\frac{\beta\Gamma[d/2-1]}
{4\pi^{d/2}\left((4\alpha^2+4\epsilon^2)\cosh^2\left(\frac{\tau+\tau'}{2\alpha}\right)\right)^{d/2-1}} \ .
\end{equation}
It may be argued by the
dominated convergence theorem that the $\epsilon$ can be dropped when calculating the 
transition rate. The geometrical reason is
that the worldline and its image under $J_B$ are totally spacelike
separated, and so the correlation function required in the
transition rate contains no divergences in the integration region.

The image term of the transition rate is thus
\begin{equation}
\label{eqn:boundaryaccel}
\dot{F}_{B\tau}(\omega)=
\frac{\beta\Gamma[d/2-1]}
{2\pi^{d/2}}
\int^{\infty}_{0}ds\,
\frac{\cos(\omega{s})}
{\left(2\alpha\cosh\left(\frac{2\tau-s}{2\alpha}\right)\right)^{d-2}} \ .
\end{equation}
We see immediately that this part of the transition rate is even in $\omega$ and hence
the boundary term breaks the KMS condition (\ref{eqn:kmscondition}). In this sense
the response is non-thermal and non-Planckian.

\begin{figure}[htbp]
\includegraphics[angle=0, width={4in}]{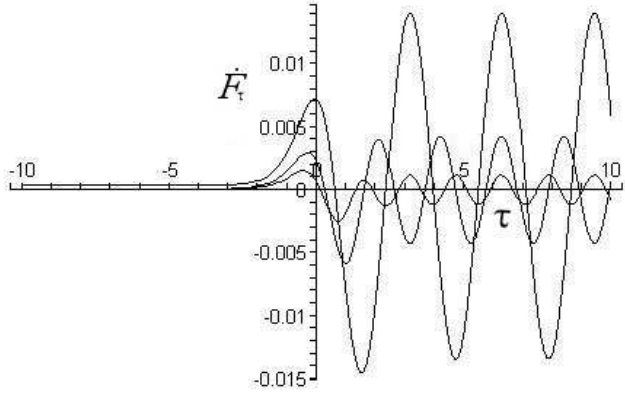}
\centering
\caption[Transition rate of uniformly accelerated detector switched on at $\tau_0=-\infty$ on $M$ with boundary, motion perpendicular to boundary.]{Transition rate for uniformly accelerated detector approaching boundary
with Neumann boundary conditions, for $\alpha=1$, $\omega=1$ (upper curve), $\omega=1.5$ and $\omega=2$ (lower curve).}
\label{fig:accel}
\end{figure}

Consider now the $4$-dimensional case, $d=4$.
When $\tau=0$ we can do the integral in (\ref{eqn:boundaryaccel})
analytically, with the result
\begin{equation}
\dot{F}_{B0}(\omega)=
\frac{\beta\omega}{4\pi\sinh(\omega\pi\alpha)} \ .
\end{equation}
For general $\tau$ we may compute the integral numerically
for different values of $\alpha$, $\tau$ and $\omega$. We have
done this with the help of Maple. For $\alpha=1$ we find the total transition rate 
(including the thermal part) displayed 
in figure \ref{fig:accel}.
Note that for many switch off times $\tau$ the image part dominates the Minkowski part.
Further we can prove analytically, by changing variables in (\ref{eqn:boundaryaccel}) by
$s=2\alpha{x}+2\tau$ expanding the $\cos$ function and evaluating the resulting integrals,
that the image part of the transition rate is given by
\begin{equation}
\label{eqn:intexpanbound}
\dot{F}_{B\tau}(\omega)=
\frac{\beta\omega\cos(2\tau\omega)}{2\pi\sinh(\omega\pi\alpha)}+B_\tau(\omega) \ ,
\end{equation}
where the function $B_\tau(\omega)$ is bounded in absolute value by $\frac{\beta}{2\pi^2\alpha}e^{-\frac{2\tau}{\alpha}}$.
Therefore for
large but finite $\tau$ the image part of the transition rate is found not to tend to
$0$ but instead is periodic in $\tau$ with period $\pi/\omega$.\footnote{For arbitrary 
dimension we may prove that 
the image part of the transition rate consists of a term periodic in $\tau$ with period $\pi/\omega$ plus a term 
bounded in absolute value by $(\mathrm{constant})e^{-\frac{(d-2)\tau}{2\alpha}}$.}
This is a property only of the transition rate of a detector which is
turned on in the infinite past. 

\begin{figure}[htbp]
\includegraphics[angle=0, width={4in}]{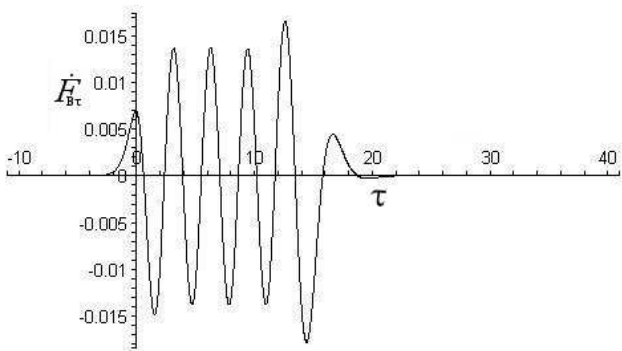}
\centering
\caption[Image term of the transition rate of a uniformly accelerated detector 
switched on at a finite time on $M$ with boundary, motion perpendicular to boundary.]
{Image term of the transition rate of a uniformly accelerated detector approaching the boundary
with Neumann boundary conditions for a detector switched on at $\tau_0=-15$, for $\alpha=1$, $\omega=1$.}
\label{fig:finitetime}
\end{figure}

Considering now a detector switched on at finite time $\tau_0$
(which recall is the more realistic situation). The
image part of the transition rate is given by (\ref{eqn:boundaryaccel})
with the upper limit of the integral replaced by $\tau-\tau_0$.
In figure \ref{fig:finitetime}
we plot this image part of the transition rate only, when the switching of the detector 
is instantaneous for, $\tau_0=-15$, $\alpha=1$ and $\omega=1$.
We find that when the detector is switched on only for a
finite time the transition rate is periodic for some time however 
falls off to the usual thermal response at late $\tau$.
We have proven this via an analytic calculation, by changing variables
in (\ref{eqn:boundaryaccel}) (with upper limit $\tau-\tau_0$) by $s=-2\alpha{y}+2\tau$ expanding the 
$\cos$ function and showing the resulting integrals are bounded in absolute value by
$Ae^{-B\tau}$ where $A$ and $B$ are postitive constants.

\begin{figure}[htbp]
\includegraphics[angle=0, width={4in}]{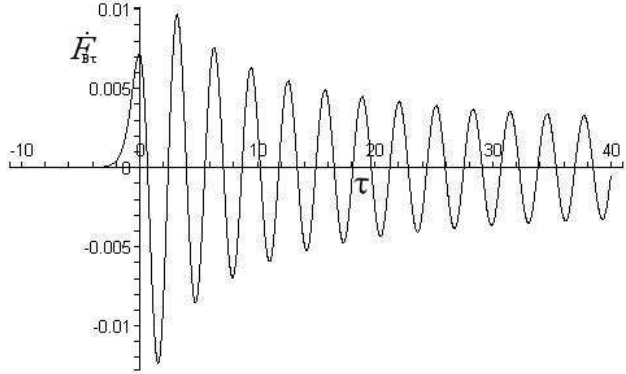}
\centering
\caption[Image term of the transition rate of a uniformly accelerated detector switched 
on at a finite time on $M$ with boundary with exponential switching function.]{Image term of the transition 
rate of a uniformly accelerated detector approaching boundary
with Neumann boundary conditions for a detector switched on at $\tau_0=-15$ with
an exponential switching function, for 
$\alpha=1$, $\omega=1$.}
\label{fig:finitetimewindow}
\end{figure}

\begin{figure}[htbp]
\includegraphics[angle=0, width={4in}]{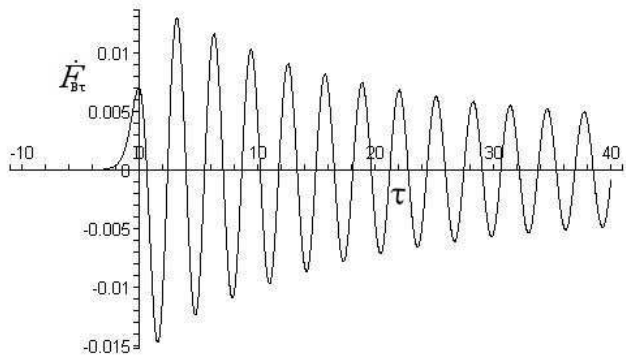}
\centering
\caption[Image term of the transition rate of a uniformly accelerated detector switched 
on at a finite time on $M$ with boundary with a Gaussian switching function.]{Image term of the transition 
rate of a uniformly accelerated detector approaching boundary
with Neumann boundary conditions for a detector switched on at $\tau_0=-15$ with
a Gaussian switching function, for 
$\alpha=1$, $\omega=1$.}
\label{fig:finitetimewindow2}
\end{figure}

In the case of instantaneous switching it was found that
even for the inertial detector in Minkowski space the response function
for a finite time detection includes a logarithmic divergence \cite{ss:svaiter}. The transition rate
however, although altered from the infinite time case, is finite for
all non-zero finite time detections.
Further in \cite{hm:higuchi} it was shown that the divergence in the 
response rate is due to the instantaneous switching: if the detector is switched on
smoothly, no divergence occurs.
It is interesting then to briefly investigate the effect of smooth switching
on the results obtained above.
We introduce therefore a smooth window function in time $\tau$ into 
the transition rate (\ref{eq:transition}), that is we consider the
rate
\begin{equation}
\label{eqn:finitetimetrans}
\dot{F}_{\tau,\tau_0}(\omega)=
2\int^{\infty}_{0}ds\,
W(s,\tau-\tau_0)
Re
\left(
e^{-i\omega{s}}\langle{0}|\phi(\tau)\phi(\tau-s)|0\rangle
\right) \ ,
\end{equation}
where $W(s,\tau-\tau_0)$ is a smooth window function
with characteristic length $\tau-\tau_0$.
In particular we consider exponential
and Gaussian switching functions
\begin{eqnarray}
\label{eqn:finitewind1}
W_1(s,\tau-\tau_0) & = & e^{-\frac{|s|}{\tau-\tau_0}} \ ,
\\
\label{eqn:finitewind2}
W_2(s,\tau-\tau_0) & = & e^{-\frac{s^2}{2(\tau-\tau_0)^2}} \ .
\end{eqnarray}
The effect of these window functions on the response
of a uniformly accelerated detector in Minkowski space
was investigated in \cite{sp:srirpad}. Here we will
only consider the effect on the image part of the
transition rate on Minkowski space with boundary.
Substituting the image term (\ref{eqn:imtermperpacc}) and one of the
window functions (\ref{eqn:finitewind1}), (\ref{eqn:finitewind2}) into the transition 
rate (\ref{eqn:finitetimetrans}), we may calculate the integral 
numerically. In figures \ref{fig:finitetimewindow} and \ref{fig:finitetimewindow2} we plot the
transition rate in four-dimensions for a sample of the parameters
and with exponential and Gaussian switching respectively.
The numerical results suggest that in all cases of finite time detection the
image part of the transition rate tends to $0$ as the detection time tends to 
infinity. That is, the transition rate tends to that on Minkowski space
in this limit, as expected as in this limit the detector recedes infinitely far
from the boundary. It is an interesting result that this was not the case
for a detector switched on in the infinite past. This suggests that in this
case the $\tau_0\rightarrow{-\infty}$ limit should be taken after the transition
rate integral has been done.

\subsection{Conical singularity and generalisations}
\label{sec:quotient}

In this section we consider the response of a uniformly accelerated
detector following trajectory (\ref{eq:accel}) on the quotient space of Minkowski space
under the group generated by the involution $J_{c_2}:(t,x,y,z)\mapsto(t,-x,-y,z)$.
Further we consider the higher dimensional generalisation of this 
spacetime constructed as the quotient of $d$-dimensional 
Minkowski space under the involution 
$J_{c_k}:(t,x_1,x_2,\ldots\,x_{d-1})\mapsto(t,-x_1,-x_2,-\ldots\,-x_k,x_{k+1},\ldots,x_{d-1})$
where $1<k<d$.
For reasons discussed later the response on these higher dimensional spaces will be
relevant when we consider the response of static detectors
on the $\rp$ geon in section \ref{sec:Geon} and inertial detectors on $\rp$ 
de Sitter space in section \ref{sec-desitrp3}. 

These spacetimes are conifolds \cite{sw:schleichwitt,sw:schleichwitt2}.
As quotients of Minkowski space under an involution with fixed points they are flat 
away from these $(d-k)$-dimensional hypersurfaces of fixed points but 
may be considered to have a distributional curvature on them
(see e.g \cite{gt:gerochtra,ga:distrbutioncurve}).
The spacetime $M/J_{c_2}$ is sometimes referred to as a conical spacetime as it 
has a conical singularity at $x_1=x_2=0$.
Transforming to cylindrical coordinates by $x_1=r\cos\phi$, $x_2=r\sin\phi$,
the isometry takes the form $J_{c_2}:(t,r,\phi,x_3,\ldots\,x_{d-1})\mapsto(t,r,\phi+\pi,x_3,\ldots,x_{d-1})$
and the metric reads
\begin{equation}
\label{cs:cosmic}
ds^2=
dt^2-dr^2-r^2d\phi^2-(dx_3)^2-\cdots-(dx_{d-1})^2 \ ,
\end{equation}
where $dr^2+r^2d\phi^2$ with the identification $(r,\phi)\sim(r,\phi+\pi)$
is the metric on a cone with deficit angle $\pi$.
In $4$ dimensions $M/J_{c_2}$  may be considered as the spacetime outside
an idealized, cosmic string \cite{vi:vilenkin} with ``gravitational mass per unit length''
$\mu=1/8$ (see \cite{an:anderson}).  

First we note that for an inertial or uniformly 
accelerated detector whose motion is in any direction $x_i$ with $k<i<d$, the 
response on these spacetimes will be the same as that
of a detector at rest or accelerating parallel to the
boundary on Minkowski space with boundary (where $\lambda$ in
(\ref{eqn:inertialparallel}) and (\ref{eqn:boundaryaccelparra2}) 
is now given by the shortest distance of the detector to the hypersurface of
fixed points $\lambda=\left((x_1)^2+(x_2)^2+\cdots+(x_k)^2\right)^{1/2}$).
This can be clearly seen by directly comparing the correlation
functions in both cases.

Consider now a particle detector uniformly accelerated with 
trajectory (\ref{eq:accel}) in the spacetime $M/J_{c_k}$.
Again as $J_{c_k}$ is an involution the correlation function
consists of two terms, 
\begin{equation}
\langle{0}|\hat{\phi}(\tau)\hat{\phi}(\tau')|0\rangle=\langle{0}|\phi(\tau)\phi(\tau')|0\rangle
+\beta\langle{0}|\phi(\tau)\phi(J_{c_k}\tau')|0\rangle \ ,
\end{equation}
where $\beta=+1, (-1)$ label the two possibilities for the
representations of the group in the automorphic field expansion 
(\ref{eqn:autofield}).
Consider first a detector switched on at $\tau_0=-\infty$.
The first term, when substituted into the transition rate (\ref{eqn:autotrans}) 
on the worldline (\ref{eq:accel}), leads to the thermal response in 
Minkowski space (\ref{thermal}).
The transition rate for the image term is
\begin{equation}
\label{eqn:boundaryaccel2}
\dot{F}_{I\tau(\omega)}=
\frac{\beta\Gamma[d/2-1]}
{2\pi^{d/2}}
\int^{\infty}_{0}ds\,
\frac{\cos(\omega{s})}
{\left(4\alpha^2\cosh^2\left(\frac{2\tau-s}{2\alpha}\right)+C_k\right)^{d/2-1}} \ ,
\end{equation}
where $C_k=\sum_{m=2}^{k}(2x_m)^2$. As the trajectory and
its image are totally spacelike separated we have dropped the
regularization in the above expression.

(\ref{eqn:boundaryaccel2}) is even in $\omega$ and so the
image term breaks the KMS condition and the response
is non-thermal and non-Planckian.
Further we may prove that (\ref{eqn:boundaryaccel2})
consists of a term periodic in $\tau$ with period $\pi/\omega$
and a term which decays
exponentially as $\tau\rightarrow\infty$. The qualitative
behaviour is therefore similar to that of the
uniformly accelerated detector on $M$ with boundary,
investigated in section \ref{sec:minkbound} (compare (\ref{eqn:boundaryaccel2}) with (\ref{eqn:boundaryaccel})).
As in section \ref{sec:minkbound} it can be shown also here that for a detector
switched on at $\tau_0>-\infty$ the image part
of the transition rate tends to $0$ as 
$\tau\rightarrow\infty$.

For a detector that accelerates towards the surface
of fixed points of the involution, that is in the
direction $r=((x_1)^2+(x_2)^2+\cdots+(x_k)^2)^{1/2}$,
the response is identical to that on Minkowski space with boundary 
with acceleration perpendicular to the boundary.
In sections \ref{sec:Geon} and \ref{sec-desitrp3}
we shall plot (\ref{eqn:boundaryaccel2}) for some 
specific values of $d$ and $k$ numerically.

We end this section with a comment on more general cosmic string spacetimes.
The methods used above could easily be applied to a larger class of idealized
cosmic string spacetimes for which the metric is (\ref{cs:cosmic}), with
the identification $(r,\phi)\sim(r,\phi+\pi/n)$ where $n\in{\Z}$
(and thus a deficit angle of $\pi(2-1/n)$), as in these
cases the correlation function may be given by a mode sum.
Detectors with motion parallel to such cosmic strings have been considered by
\cite{ss:cosmic,ds:cosdet}. Their main conclusions, which agree with ours
here where they overlap, are that the detector does respond to the presence of
the string in a manner which depends on its distance from the string.
Our results above and in the previous subsection add to the discussion, as we have been 
able to show the behaviour of detectors when motion is perpendicular to the string for
the specific case of $n=1$.
Numerical evaluations of the transition rate for any $n$ could be done in a similar 
way, but we shall not pursue this further here.
It is important to note however this class of cosmic strings does not include
realistic cosmic strings of the GUT scale where the deficit angle is $\approx{10^{-5}}$.

\subsection{Scalar detector on $M_-$}

Finally let us consider $M_-$ \cite{lm:geon,pl:langlois,pl:langlois2} which is 
a quotient of Minkowski space (or of $M_0$, it being a double cover of $M_-$) under the map
$J_-:(t,x,y,z)\mapsto(t,-x,-y,z+a)$. Our interest in $M_-$, as well as it being an interesting 
topologically non-trivial spacetime in which we can probe the effect of topology on the Unruh effect,
lies in its role in modelling, via accelerated observers on 
flat spacetimes, the Hawking(-Unruh) effect on the $\mathbb{RP}^3$ geon \cite{lm:geon}.

Again we may use the method of images
to find the correlation function for the automorphic field, with the result
\begin{equation}
\label{eqn:m-imagesum}
\langle{0}|\hat{\phi}(\tau)\hat{\phi}(\tau')|0\rangle=\sum_{n\in{\Z}}\langle{0}|\phi(\tau)\phi(J_-\tau')|0\rangle \ .
\end{equation}
The transition rate is given by (\ref{eqn:autotrans}).

\subsubsection{Inertial detector}

Considering a detector following the inertial trajectory
(\ref{eqn:inertraj}). The transition rate may again be split into two parts. The first comes from 
the $M_0$ part in the image sum (the even $n$ terms in (\ref{eqn:m-imagesum})) and will lead to
the same response as on $M_0$ for untwisted fields (\ref{eqn:inerm0}). The other part (due to odd $n$
terms in (\ref{eqn:m-imagesum})) is similar
to the boundary part of such a detector on $M$ with boundary,
giving in the transition rate the contribution
\begin{eqnarray}
\dot{F}_{I\tau(\omega)} & = & 
-\frac{1}{2\pi^2}
(1-v^2)
\sum_{n=-\infty}^{\infty}
\frac{1}
{C_n+B_n}
(
-\mathrm{Ci}(B_n|\omega|)\cos(B_n|\omega|)
\nonumber
\\
&  & -\mathrm{si}(B_n|\omega|)\sin(B_n|\omega|)
+\mathrm{Ci}(C_n|\omega|)\cos(C_n|\omega|)
+\mathrm{si}(C_n|\omega|)\sin(C_n|\omega|)
\nonumber
\\
&  & +2\pi\sin(B_n\omega)\Theta(-\omega)
) \ ,
\end{eqnarray}
where
\begin{eqnarray}
B_n=\frac{-4\tau{v}^2+(16\tau^2{v^2}+4(1-v^2)^2((2y_0)^2+(2na+a)^2))^{1/2}}
{2(1-v^2)} \ ,
\\
C_n=\frac{+4\tau{v^2}+(16\tau^2{v^2}+4(1-v^2)^2((2y_0)^2+(2na+a)^2))^{1/2}}
{2(1-v^2)} \ .
\end{eqnarray}

In contrast to the analogous result on $M$ with boundary, there is no divergence
here at $x=0$ as there is no obstruction there and the inertial detector on
$M_-$ carries through $x=0$ smoothly. Note that on $M_-$ the energy-momentum
tensor expectation 
values are finite over the whole spacetime \cite{lm:geon,pl:langlois2}, while on $M$ with
boundary they diverge at $x=0$. For $\omega>0$ the $M_0$ part of the transition rate
vanishes while the image part is odd in $\tau$.
Further, numerical evaluations of the sum indicate that it is non-zero.
Here therefore we have an example of a spacetime and trajectory with no
pathologies at all where the total transition rate is negative for some values
of proper time $\tau$.

\subsubsection{Uniformly accelerated detector}

Consider a detector following the uniformly accelerated
worldline (\ref{eq:accel}).
Again the correlation function is in two parts. The part coming from
the $M_0$ part of the image sum  is given by the corresponding response on $M_0$
((\ref{M0response}) with $\eta=-1$). This part satisfies the KMS condition
and so in this sense is thermal. The ``image'' part 
is then somewhat similar to the boundary part found in the case
of $M$ with boundary. The transition rate for this term is 
\begin{equation}
\label{eqn:m-accelbound}
\dot{F}_{I\tau}(\omega)=
\frac{1}{2\pi^2}
\sum_{n=-\infty}^\infty
\int^{\infty}_{0}ds\,
\frac{\cos(\omega{s})}
{4\alpha^2\cosh^2\left({\frac{2\tau-s}{2\alpha}}\right)+4y_0^2+a^2(2n-1)^2} \ ,
\end{equation}
where $y_0$ is the $y$ coordinate of the detector. 
As on Minkowski space with boundary this image part of the transition rate
is even in $\omega$ and so breaks the KMS condition. The transition rate
is thus non-thermal and non-Planckian.
Further we find, by similar techniques to those leading to (\ref{eqn:intexpanbound})
\begin{equation}
\dot{F}_{I\tau}(\omega)=
\sum_{n=-\infty}^\infty
\left(
\frac{\alpha\cos(2\tau\omega)\sin\left(\alpha\omega{\mathrm{arccosh}\left(\frac{c_n}{2\alpha^2}\right)}\right)}
{\pi(c_n^2-4\alpha^4)^{1/2}\sinh(\omega\pi\alpha)}
+B_{n,\tau}(\omega)
\right) \ ,
\end{equation}
where $c_n=2\alpha^2+4y_0^2+(2na-a)^2$, and each $B_{n,\tau}(\omega)$ is bounded
by $1/{(2\pi^2\alpha)}e^{-2\tau/\alpha}$. Further we may show
that the sum of $B_{n,\tau}(\omega)$ over $n$ is bounded by a function which exponentially
decays as $\tau\rightarrow\infty$.
We see as with the detector on $M$ with
boundary for large $\tau$ the response becomes periodic in $\tau$ with period $\pi/\omega$.

We may investigate the
general case here numerically. An analytic result is easy to find in the
case when $\tau=0$, with the result
\begin{equation}
\dot{F}_{I0}(\omega)=
\frac{\alpha}{2\pi}
\sum_{n=-\infty}^\infty
\frac{\sin\left(\alpha\omega{\mathrm{arccosh}\left(\frac{c_n}{2\alpha^2}\right)}\right)}
{(c_n^2-4\alpha^4)^{1/2}\sinh(\omega\pi\alpha)}  \ .
\end{equation}

For a detector switched on instantaneously at a finite time $\tau_0>-\infty$ 
an analytic calculation shows that the difference
between the response on $M_-$ and that on $M_0$ dies off as $\tau\rightarrow\infty$,
as would be expected far away from $x=0$. It is an interesting point that this is 
not the case for the detector switched on the infinite past.

This clarifies and adds to the
discussion on particle detectors given in \cite{lm:geon}.

\section{Causal detector for the Dirac field}
\label{sec:dirac}

\subsection{Minkowski space}
\label{sec:diracmink}

In this section we extend the causal detector to the massless Dirac field in four-dimensional
Minkowski space. The detector is
still a many-level quantum mechanical system with free Hamiltonian
$H_D$. However now the detector moves through a massless Dirac field
$\psi$ (with free Hamiltonian $H_\psi$) in Minkowski space to which it 
is coupled via the interaction Hamiltonian
\begin{equation}
\label{eqn:dirachint}
H_{\mathrm{int}}=
c
m(\tau)
\bar{\psi}(\tau)
\psi(\tau) \ ,
\end{equation}
where $\bar{\psi}=\psi^\dagger\gamma^0$, and $\psi(\tau)=\psi(x(\tau))$.
The equation of motion for the free field $\psi$ is the massless free
Dirac equation $i\gamma^\mu\partial_\mu\psi=0$. We choose a basis of solutions and expand the
field in terms of this basis. We work throughout with the standard representation
of $\gamma$ matrices,
\begin{eqnarray} \gamma^0=\left(\begin{array}{cc} 1 & 0 \\
		 0 & -1 \end{array}\right) \ , & \gamma^i=\left(\begin{array}{cc} 0 & \sigma_i \\
		-\sigma_i & 0 \end{array}\right) \ ,
\end{eqnarray}
where $\sigma_i$ are the Pauli matrices, $\sigma_1 =
\bigl(
\begin{smallmatrix} 0 & 1 \\ 1 & 0
\end{smallmatrix} 
\bigr)$,  
$\sigma_2 = 
\bigl(
\begin{smallmatrix} 
0 & -i \\ i & 0 
\end{smallmatrix}
\bigr)$
and 
$\sigma_3 = 
\bigl(
\begin{smallmatrix} 
1 & 0 \\ 0 & -1
\end{smallmatrix}
\bigr)$. Then
\begin{equation}
\label{eqn:diracexpanmode}
\psi(t,\mathbf{x})
=
\sum_{s=1,2}
\int
\frac{d^3k}{(2(2\pi)^3)^{\frac{1}{2}}}
\left[
b_s(\mathbf{k})u(\mathbf{k},s)e^{-i\omega{t}+i\mathbf{k}\mathbf{x}}+
d^\dagger_s(\mathbf{k})v(\mathbf{k},s)e^{i\omega{t}-i\mathbf{k}\mathbf{x}}
\right] \ ,
\end{equation}
where
\begin{eqnarray}
\label{eqn:uspinors}
u(\mathbf{k},1)=\left(\begin{array}{ccccccccccccccccccccc}1 \\ 0 \\ \frac{k_z}{\omega} \\ \frac{k_+}{\omega} \end{array} \right) \ ,
 &       \hspace{.25in}           u(\mathbf{k},2)=\left(\begin{array}{ccccccccccccccccccccc}0 \\ 1 \\ \frac{k_-}{\omega}\\ \frac{-k_z}{\omega} \end{array} \right) \ ,
\end{eqnarray}	
and
\begin{eqnarray}
v(\mathbf{k},1)=\left(\begin{array}{ccccccccccccccccccccc}\frac{k_z}{\omega} \\ \frac{k_+}{\omega} \\ 1 \\ 0 \end{array} \right) \ ,
 &       \hspace{.25in}
v(\mathbf{k},2)=\left(\begin{array}{ccccccccccccccccccccc}\frac{k_-}{\omega} \\ \frac{-k_z}{\omega} \\ 0 \\ 1 \end{array} \right) \ ,
\end{eqnarray}	
with $k_{\pm}=k_x\pm{ik_y}$. The modes in the expansion (\ref{eqn:diracexpanmode}) are expressed in
terms of a standard Minkowski vierbein, aligned along Minkowski coordinate axes,
and they are suitably 
orthonormal with respect to the usual Dirac inner product in Minkowski 
space,
\begin{equation}
\langle{\psi_1,\psi_2}\rangle=\int{d^3x}\,\psi_1^\dagger{\psi_2} \ .
\end{equation}
The free field is then quantized in the usual manner, imposing the
usual anticommutation relations on the annihilation/creation
operators.

We assume again that at time $\tau_0$ the full interacting field
is in the product state $|0,E_0\rangle=|0\rangle|E_0\rangle$. 
Working in the interaction picture we find, to first order in perturbation theory,
that the probability that at a later time $\tau_1>\tau_0$ the detector is found 
in state $|E_1\rangle$ is
\begin{eqnarray}
\sum_{\Psi}
|\langle\Psi,E_1|0,E_0\rangle|^2
& = & 
c^2
|\langle{E_1}|m(0)|E_0\rangle|^2
\int^{\tau_1}_{\tau_0}d\tau
\int^{{\tau}_1}_{{\tau}_0}d\tau'
e^{-i\omega(\tau-\tau')}
\nonumber
\\
&   &
\times
\langle{0}|\bar{\psi}(\tau)\psi(\tau)\bar{\psi}(\tau')\psi(\tau')|0\rangle
 \ ,
\end{eqnarray}
with $\omega=E_1-E_0$. Once again we shall concentrate on the response function part.
With the same change of coordinates as in section \ref{sec-linmink}, and differentiating with
respect to $\tau_1=\tau$
we obtain the transition rate,
\begin{equation}
\label{eq:transitiondir}
\dot{F}_\tau(\omega)=
2\int^{\infty}_{0}ds\,
Re
\left(
e^{-i\omega{s}}\langle{0}|\bar{\psi}(\tau)\psi(\tau)\bar{\psi}(\tau-s)\psi(\tau-s)|0\rangle
\right) \ .
\end{equation}
Further here we find
\begin{equation}
\label{eq:diraccorr}
\langle{0}|\bar{\psi}(\tau)\psi(\tau)\bar{\psi}(\tau')\psi(\tau')|0\rangle
=
\mathrm{Tr}((S^+_M(\tau,\tau'))^2) \ ,
\end{equation}
where $\mathrm{Tr}$ is the trace and $S^+_M(\tau,\tau')=\langle{0}|\psi(\tau)\bar{\psi}(\tau')|0\rangle$ 
is the positive frequency Wightman function, which is related to the scalar field 
positive frequency Wightman function by (see e.g \cite{bd:book})
\begin{equation}
S^+_M(\tau,\tau')=
i\gamma^{\mu}\partial_{\mu}G^+_M(\tau,\tau') \ .
\end{equation}
We note here that all expressions for the response are independent of the vierbein used
to express the $\psi$ field. This is due to the form of $H_{\mathrm{int}}$ (\ref{eqn:dirachint}) 
which is a Lorentz scalar.
We also note that in the case of a massive Dirac field in Minkowski space (\ref{eq:diraccorr}) contains a second
term proportional to $\mathrm{Tr}(S^+_M(\tau,\tau'))\mathrm{Tr}(S^-_M(\tau',\tau))$. 
Here in the massless case this term does not enter as $\mathrm{Tr}(S^+_M(\tau,\tau'))=0$
for any worldline.

Consider the uniformly accelerated worldline
(\ref{eq:accel}).
Again a numerical calculation shows that if we use, in the scalar field correlation
function above, the $i\epsilon$ regularization we will 
get a $\tau$-dependent result for the transition rate (\ref{eq:transitiondir}) even in the $\epsilon\rightarrow{0}$ limit.
We are thus led once again to consider an alternative regularization
where we use a smeared form for the field operator in the interaction
Hamiltonian. That is we consider
\begin{equation}
\label{eq:smeardir}
\psi(\tau)=
\int
d^3\xi\,
W_\epsilon(\mathbf{\xi})
S(\tau,\mathbf{\xi})
\psi(x(\tau,\mathbf{\xi})) \ ,
\end{equation}
with the same definitions for $W_\epsilon(\mathbf{\xi})$ and $\mathbf{\xi}$
as in section \ref{sec-linmink}. In contrast to the scalar case, here
we include $S(\tau,\mathbf{\xi})$, which is the spinor transformation associated with the transformation 
from the Minkowski vierbein
to one adapted to the Fermi-Walker coordinates. However we may now argue that in this case
$S(\tau,\mathbf{\xi})$ may be dropped. Firstly we note that the metric written in 
Fermi Walker coordinates is \cite{sc:schlicht}
\begin{equation}
ds^2=
\left(1+2\left(\dot{t}\ddot{x}-\dot{x}\ddot{t}\right)\xi+\left(\ddot{x}^2-\ddot{t}^2\right)\xi^2\right)
d\tau^2
-
d\mathbf{\xi}^2 \ .
\end{equation}
Constant $\tau$ spatial sections are therefore flat. It then follows
that the transformation from Minkowski vierbein to that adapted to
these Fermi coordinates will be independent of $\mathbf{\xi}$, as Fermi Walker
transport along these spatial sections in a non-rotating vierbein will
be trivial. It therefore follows that $S(\tau,\mathbf{\xi})$ may be taken outside
the integral in (\ref{eq:smeardir}). Further it follows from the form of $H_{\mathrm{int}}$
(\ref{eqn:dirachint}) that as $S$ may be taken outside the integral it may be dropped completely.
Therefore on $M$ we may work throughout with expressions written with respect
to the standard Minkowski vierbein and with 
\begin{equation}
\label{eqn:smeardiracfi}
\psi(\tau)=
\int
d^3\xi\,
W_\epsilon(\mathbf{\xi})
\psi(x(\tau,\mathbf{\xi})) \ ,
\end{equation}
as the expression for a smeared field operator.
By arguments similar to those that lead to (\ref{eq:transitiondir}), 
the transition rate is 
given by 
\begin{equation}
\label{eqn:smeartransdir}
\dot{F}_\tau(\omega)=
2\int^{\infty}_{0}ds\,
Re
\left(
e^{-i\omega{s}}\langle{0}|\bar{\psi}(\tau)\psi(\tau)\bar{\psi}(\tau-s)\psi(\tau-s)|0\rangle
\right) \ ,
\end{equation}
and we find
\begin{equation}
\label{eqn:smearcorrdirac}
\langle{0}|\bar{\psi}(\tau)\psi(\tau)\bar{\psi}(\tau')\psi(\tau')|0\rangle
=
\mathrm{Tr}(\langle{0}|\psi(\tau)\bar{\psi}(\tau')|0\rangle^2) \ ,
\end{equation}
where $\psi(\tau)$ is now the smeared field (\ref{eqn:smeardiracfi}).

Now suppose we consider again the Lorentzian profile function
\begin{equation}
W_{\epsilon}(\mathbf{\xi})
=
\frac{1}{\pi^{2}}
\frac{\epsilon}{(\mathbf{\xi}^2+\epsilon^2)^{2}} \ .
\end{equation}
The spinor correlation function is then given by
\begin{eqnarray}
\langle{0}|\psi(\tau)\bar{\psi}(\tau')|0\rangle & = & 
\frac{1}{2(2\pi)^3}
\sum_{s=1,2}
\int
d^3k\,
u(\mathbf{k},s)
u^\dagger(\mathbf{k},s)
\gamma^0
\nonumber
\\
&  & \times
\int
d^3\xi\,
W_\epsilon(\mathbf{\xi})
e^{-ik\cdot{x}(\tau,\mathbf{\xi})}
\int
d^3\xi'\,
W_\epsilon(\mathbf{\xi'})
e^{ik\cdot{x}(\tau,\mathbf{\xi'})} \ .
\label{eqn:diraccorrint}
\end{eqnarray}
The integrals over $\mathbf{\xi}$ and $\mathbf{\xi}'$ in (\ref{eqn:diraccorrint})
are the same as those in (\ref{eq:expan}). Proceeding as with (\ref{eq:expan}), we find
\begin{equation}
\label{eqn:dircorrinteq}
\langle{0}|\psi(\tau)\bar{\psi}(\tau')|0\rangle=
\frac{1}{2(2\pi)^3}
\sum_{s=1,2}
\int
d^3k\,
u(\mathbf{k},s)
u^\dagger(\mathbf{k},s)
\gamma^0
e^{-i\omega(t-t'-i\epsilon(\dot{t}+\dot{t}'))+i\mathbf{k}(\mathbf{x}-\mathbf{x'}-i\epsilon(\dot{\mathbf{x}}+\dot{\mathbf{x}}'))}
 \ .
\end{equation}
Comparing (\ref{eqn:dircorrinteq}) with the scalar field expression
(\ref{eq:correlation}), with $d=4$ and using (\ref{eqn:uspinors}), we find we have
\begin{equation}
\label{eq:SeqG}
S^+_M(\tau,\tau')=\langle{0}|\psi(\tau)\bar{\psi}(\tau')|0\rangle
=
i\gamma^\mu\partial_\mu{\langle{0}|\phi(\tau)\bar{\phi}(\tau')|0\rangle} \ ,
\end{equation}
where $\langle{0}|\phi(\tau)\bar{\phi}(\tau')|0\rangle$ is
the scalar field correlation function (\ref{eq:correlation})
and the partial derivative acts on the $(t,\mathbf{x})$
but NOT on the $(\dot{t},\dot{\mathbf{x}})$. Therefore
we find, from
(\ref{eq:corre}), that the spinor correlation function is given by
\begin{eqnarray}
\label{eqn:dirccorrfnsmeared}
S^+_M(\tau,\tau') & = & \frac{i}{4{\pi}^2}\frac{1}{[\tilde{t}^2-\tilde{x}^2-\tilde{y}^2-\tilde{z}^2]^2}
\nonumber
\\
&  &
\times
\left[
\begin{array}{cccc} 
2\tilde{t} & 0 & -2\tilde{z} & 2(i\tilde{y}-\tilde{x}) 
\\ 
0 & 2\tilde{t} & -2(i\tilde{y}+\tilde{x}) & 2\tilde{z} 
\\
2\tilde{z} & -2(i\tilde{y}-\tilde{x}) & 2\tilde{t} & 0 
\\
2(i\tilde{y}+\tilde{x}) & -2\tilde{z} & 0 & 2\tilde{t}
\end{array} 
\right] \ ,
\end{eqnarray}
where $\tilde{a}=(a(\tau)-a(\tau')-i\epsilon(\dot{a}(\tau)+\dot{a}(\tau')))$.
From this it is easy to show that
\begin{equation}
\label{eqn:diraccorrefunction}
\mathrm{Tr}(S^+_M(\tau,\tau')^2)
=
-\frac{1}{\pi^4}
\frac{1}{((t-t'-i\epsilon(\dot{t}+\dot{t}'))^2-(\mathbf{x}-\mathbf{x'}-i\epsilon(\dot{\mathbf{x}}+\dot{\mathbf{x}}'))^2)^3}
 \ .
\end{equation}

\subsubsection{Inertial detector}

First we consider the response of a Dirac field detector following the
inertial worldline (\ref{eqn:inertraj}) in Minkowski space. From
(\ref{eqn:diraccorrefunction}), (\ref{eqn:smeartransdir}) and (\ref{eqn:smearcorrdirac})
the transition rate is found to be
\begin{equation}
\label{eqn:diracinertmink}
\dot{F}_\tau(\omega)=
-\frac{1}{\pi^4}
\lim_{\epsilon\rightarrow{0}}
\int^{\infty}_{-\infty}ds\,
\frac{e^{-i\omega{s}}}{(s-2i\epsilon)^6} \ .
\end{equation}
The integral can be done by residues, with the result
\begin{equation}
\label{eqn:diracinertitrans}
\dot{F}_\tau(\omega)=
-\frac{\omega^5}{60\pi^3}\Theta(-\omega) \ .
\end{equation}

Consider also the ``power spectrum'' of the Dirac noise
as defined by Takagi in \cite{tk:takagi}. The noise $g(\tau,\tau')$
is defined by
\begin{equation}
g(\tau,\tau')
=
S(\tau)
S^+_M(\tau,\tau')
S({\tau'})^{-1} \ ,
\end{equation}
where $S(\tau)=S(\tau,\mathbf{\xi})$ as given in (\ref{eq:smeardir}). $S(\tau)$
is the spinor transformation which takes care of the Fermi-Walker transport
so that $S(\tau)\psi(\tau)$ does not rotate with respect to the detector's
proper reference frame.
The definition for the power spectrum on a stationary worldline,
where $S^+_M(\tau,\tau')$ depends on $\tau$ and $\tau'$ only through
$\tau-\tau'$, is
\begin{equation}
\label{eq:powerspec}
P(\omega)=\frac{1}{4}
\mathrm{Tr}\,\gamma^0
\int^{\infty}_{-\infty}
d\tau\,
{e^{-i\omega\tau}}g(\tau) \ .
\end{equation}
On the inertial trajectory (\ref{eqn:inertraj}), the transformation to the
Fermi frame is trivial and we find
\begin{eqnarray}
P(\omega) & = & \frac{i}{2\pi^2}
\int^{\infty}_{-\infty}
ds\,
\frac{e^{-i\omega{s}}}{(s-2i\epsilon)^3}
\nonumber
\\
& = & \frac{\omega^2}{2\pi}\Theta(-\omega) \ .
\end{eqnarray}
We note that the power spectrum is $-\omega$ times the transition
rate for the linearly coupled scalar field detector 
following the same trajectory.

\subsubsection{Uniformly accelerated detector}

Considering once again a detector following the uniformly accelerated worldline (\ref{eq:accel}).
We find, as expected, that the correlation function is invariant under translations in $\tau$ and the 
transition rate (\ref{eqn:smeartransdir}) is given by
\begin{equation}
\dot{F}_\tau(\omega)=
-\frac{1}{64\pi^4}
\lim_{\epsilon\rightarrow{0}}
\int^{\infty}_{-\infty}ds\,
\frac{e^{-i\omega{s}}}{\left(\alpha\sinh\left(\frac{s}{2\alpha}\right)-i\epsilon\cosh\left(\frac{s}{2\alpha}\right)\right)^6} \ .
\end{equation}
The integral may be done by contour integration. The result is
\begin{equation}
\label{eq:diractranm}
\dot{F}_\tau(\omega)=
\frac{1}{60\pi^3\alpha^4}
\frac{\omega}{\left(e^{2\pi\alpha\omega}-1\right)}
(4+5(\alpha\omega)^2+(\alpha\omega)^4) \ .
\end{equation}
The response is thermal in the sense that it satisfies the KMS
condition at the temperature $T=(2\pi\alpha)^{-1}$. It is interesting to note that there is no fermionic 
factor in the response, instead we have the usual Planckian factor 
found in the scalar case. 

We can see the fermionic factor appearing
however if we consider the power spectrum (\ref{eq:powerspec}) of the Dirac noise.
For the uniformly accelerated worldline we have
\begin{equation}
S(\tau)
=
\cosh\left(\frac{\tau}{2\alpha}\right)-\gamma^0\gamma^1\sinh\left(\frac{\tau}{2\alpha}\right) \ ,
\end{equation}
and the power spectrum (\ref{eq:powerspec}) is given by
\begin{equation}
P(\omega)=\frac{i}{16\pi^2}
\int^{\infty}_{-\infty}
d\tau\,
\frac{e^{-i\omega\tau}}{\left(\alpha\sinh\left(\frac{\tau}{2\alpha}\right)-i\epsilon\cosh\left(\frac{\tau}{2\alpha}\right)\right)^3} \ .
\end{equation}
Again the integral may be done by contour integration to give
\begin{equation}
\label{eqn:minkdiracpspec}
P(\omega)=
\frac{1}{2\pi}
\frac{\left(\omega^2+\frac{1}{4\alpha^2}\right)}{\left(1+e^{2\pi\alpha\omega}\right)} \ .
\end{equation}
Our expression here agrees with that found by Takagi \cite{tk:takagi}.

\subsection{Dirac detector for automorphic fields}
\label{sec:diracauto}

Next we wish to consider this fermionic detector on $M_0$ and
$M_-$ and in particular address the issues concerning spin structure
on $M_-$ \cite{pl:langlois}. 
We consider an automorphic Dirac field on Minkowski space. The main
difference from the scalar case is that we must take care of what vierbeins
our expressions are written with respect to. In particular our vierbein
might not be invariant under the quotient group $\Gamma$. 

We begin with a massless 
Dirac field $\psi$
on $M$, expressed with respect to a vierbein
that is invariant under $\Gamma$. The automorphic field is then defined by
\begin{equation}
\label{eqn:diracautofield}
\hat{\psi}(x)=
\frac{1}
{\left(\sum_{\gamma\in\Gamma}p(\gamma)^2\right)^{1/2}}
\sum_{\gamma\in\Gamma}
p(\gamma)
\psi
(\gamma^{-1}x) \ ,
\end{equation}
where the normalization is such that, at equal times
\begin{equation}
\left\{\hat{\psi}_\alpha(x),\hat{\psi}^\dagger_\beta(x')\right\}=\delta^{(d-1)}(x-x')\delta_{\alpha\beta} + \textrm{image terms} \ .
\end{equation}
The two-point function for the automorphic field is then given by the method of images,
\begin{equation}
\label{eq:diracautocorr}
S^+_{M/\Gamma}(x,x')=
\langle{0}|\hat{\psi}(x)\bar{\hat{\psi}}(x')|0\rangle
=
\sum_{\gamma\in\Gamma}
p(\gamma)
\langle{0}|\psi(x)\bar{\psi}(\gamma^{-1}x')|0\rangle \ .
\end{equation}

We consider a detector coupled to the automorphic field
via the interaction Hamiltonian
\begin{equation}
H_{\mathrm{int}}
=
cm(\tau)
\bar{\hat{\psi}}(\tau)\hat{\psi}(\tau) \ ,
\end{equation}
where
\begin{equation}
\label{eqn:smaereddirac}
\hat{\psi}(\tau)=
\int
d^3\xi
\,
W_\epsilon(\mathbf{\xi})
S(\tau,\mathbf{\xi})
\hat{\psi}(x(\tau,\mathbf{\xi})) \ .
\end{equation}
The transition rate is given by
\begin{equation}
\dot{F}_\tau(\omega)=
2\int^{\infty}_{0}ds\,
Re
\left(
e^{-i\omega{s}}\langle{0}|\bar{\hat{\psi}}(\tau)\hat{\psi}(\tau)\bar{\hat{\psi}}(\tau-s)\hat{\psi}(\tau-s)|0\rangle
\right) \ .
\end{equation}
Further we may show, with a calculation similar to that leading to (\ref{eq:diraccorr}) and (\ref{eqn:smearcorrdirac}), that
\begin{equation}
\langle{0}|\bar{\hat{\psi}}(\tau)\hat{\psi}(\tau)\bar{\hat{\psi}}(\tau')\hat{\psi}(\tau')|0\rangle
=
\mathrm{Tr}
\left(\langle{0}|\hat{\psi}(\tau)\bar{\hat{\psi}}(\tau')|0\rangle^2\right) \ ,
\end{equation}
with
\begin{equation}
\label{eqn:autodircorrfn}
\langle{0}|\hat{\psi}(\tau)\bar{\hat{\psi}}(\tau')|0\rangle
=
\sum_{\gamma\in\Gamma}
p(\gamma)
\langle{0}|\psi(\tau)\bar{\psi}(\gamma^{-1}\tau')|0\rangle \ .
\end{equation}
Therefore the method of images may be directly applied to our 
Minkowski space correlation functions here. 

It is important to note that the above mode sum expressions are changed when considering
a vierbein not invariant under the action of $\Gamma$. Suppose we
consider two vierbeins, one invariant under $\Gamma$ (labelled by
an $I$) and another not invariant (labelled by $N$). In the
vierbein $I$ the automorphic field is given by the mode sum expression
(\ref{eqn:diracautofield}). The transformation from $I$ to $N$ will
transform the spinors as $\hat{\psi}_I(x)\rightarrow\hat{\psi}_N(x)=S(x)\hat{\psi}_I(x)$.
Then from (\ref{eqn:diracautofield})
\begin{equation}
\hat{\psi}_N(x)=
\frac{1}
{\left(\sum_{\gamma\in\Gamma}p(\gamma)^2\right)^{1/2}}
\sum_{\gamma\in\Gamma}
p(\gamma)
S(x)
\psi_I
(\gamma^{-1}x) \ ,
\end{equation}
and hence the mode sum expression for the automorphic field in terms
of the non-invariant vierbein is
\begin{equation}
\hat{\psi}_N(x)=
\frac{1}
{\left(\sum_{\gamma\in\Gamma}p(\gamma)^2\right)^{1/2}}
\sum_{\gamma\in\Gamma}
p(\gamma)
S(x)S^{-1}(\gamma^{-1}x)
\psi_N
(\gamma^{-1}x) \ .
\end{equation}

Similarly the two-point function tranforms as 
\begin{equation}
S^+_{IM/\Gamma}(x,x')
\rightarrow{S^+_{NM/\Gamma}(x,x')}=S(x)S^+_{IM/\Gamma}(x,x')S^{-1}(x') \ .
\end{equation}
From (\ref{eq:diracautocorr}), the mode sum expression for the two point function
in terms of the non-invariant vierbein is hence
\begin{eqnarray}
S^+_{NM/\Gamma}(x,x')
& = & 
\sum_{\gamma\in\Gamma}
p(\gamma)
S(x)\langle{0}|\psi_I(x)\bar{\psi}_I(\gamma^{-1}x')|0\rangle{S^{-1}(x')} \ ,
\nonumber
\\
& = & 
\sum_{\gamma\in\Gamma}
p(\gamma)
\langle{0}|\psi_N(x)\bar{\psi}_N(\gamma^{-1}x')|0\rangle{S^{-1}(\gamma^{-1}x')}S^{-1}(x') \ .
\end{eqnarray}

In sections \ref{sec:diracm0} and \ref{sec-diracM-} we
shall work throughout in vierbeins invariant under
$J_0$ and $J_-$ respectively.

\subsection{Dirac detector on $M_0$}
\label{sec:diracm0}

Consider now Dirac field theory on $M_0$ as an automorphic field theory
on $M$ where expressions are written with respect to the standard Minkowski 
vierbein. From (\ref{eqn:autodircorrfn}) the $M_0$ correlation function is 
given by
\begin{equation}
\label{eq:diraccorM0}
S^+_{M_0}(\tau,\tau')=
\sum_{n\in\Z}
\eta^n
S^+_{M}(\tau,J_0(\tau')) \ ,
\end{equation}
where $\eta=1,(-1)$ labels spinors with periodic (antiperiodic)
boundary conditions. We therefore find an explicit expression
for $S^+_{M_0}(\tau,\tau')$ from (\ref{eq:diraccorM0}) and (\ref{eqn:dirccorrfnsmeared}).
As we work throughout in the standard Minkowski vierbein here,
writing the smeared field operator as in (\ref{eqn:smaereddirac}),
we may again argue in an analogous way to in the
previous section that the spinor transformation
$S(\tau,\mathbf{\xi})$ can be dropped.

\subsubsection{Inertial detector}

First consider a Dirac detector following the inertial worldline
(\ref{eqn:inertraj}) on $M_0$.
The power spectrum (\ref{eq:powerspec}) for the noise is found to be
\begin{equation}
P(\omega)=\left(\frac{\omega^2}{2\pi}
+2
\sum_{n=1}^{\infty}
\frac{\eta^n\omega}
{4\pi{n}a}
\sin(2n\omega{a})\right)
\Theta(-\omega) \ .
\end{equation}
As on Minkowski space the power spectrum is $-\omega$ 
times the transition rate of the linearly coupled scalar field detector
following the same trajectory (\ref{eqn:transm0scalarinert}).
The summation thus may be performed to give $-\omega$ times
(\ref{eqn:inerm0}) and (\ref{eqn:inerm2}). 

For the transition rate we find
\begin{equation}
\dot{F}_\tau(\omega)
=
-\frac{1}{\pi^4}
\sum_{n\in\Z}
\sum_{m\in\Z}
\int_{-\infty}^{\infty}
ds\,
\frac{\eta^n\eta^m((s-2i\epsilon)^2-4nma^2)}
{[(s-2i\epsilon)^2-(2na)^2]^2[(s-2i\epsilon)^2-(2ma)^2]^2} \ .
\end{equation}
The $n=0,m=0$ term gives the transition rate on Minkowski space
(\ref{eqn:diracinertitrans}). The integral for other terms may 
be done by residues, with the result
\begin{eqnarray}
\dot{F}_\tau(\omega) & = & 
\Bigg(
-\frac{\omega^5}{60\pi^3}
+
\frac{1}{32\pi}
\sum_{\substack{n.m=-\infty \\ n,m\neq{0}}}^{\infty}
\frac{\eta^n\eta^m}
{(m-n)(m+n)^3a^5}
\left[
\left(\left(\frac{2m\omega{a}}{n}+2\omega{a}\right)\cos(2\omega{n}a)
\right.
\right.
\nonumber
\\
& &\left.-\frac{(m+3n)}{n^2}\sin(2\omega{n}{a})\right)
+
\left(\left(\frac{2\omega{a}n}{m}+2\omega{a}\right)\cos(2\omega{m}a)
\right.
\nonumber
\\
& &
\left.
\left.
-\frac{(n+3m)}{m^2}\sin(2\omega{m}{a})\right)\right]\Bigg)
\Theta(-\omega) \ ,
\end{eqnarray}
where the $n=m$ and $n=-m$ terms are understood in the limiting sense
and can be verified to be finite.
As expected the response does not depend on the velocity.

\subsubsection{Uniformly accelerated detector}

Next consider the power spectrum for the Rindler noise, that is 
we consider $g(\tau,\tau')$ on the uniformly accelerated worldline.
From (\ref{eq:powerspec}) we find 
\begin{equation}
P(\omega)=\frac{i}{16\pi^2}
\int^{\infty}_{-\infty}
d\tau
\sum_{n\in\Z}
\frac{\eta^n{e^{-i\omega\tau}}\left(\alpha\sinh\left(\frac{\tau}{2\alpha}\right)-i\epsilon\cosh\left(\frac{\tau}{2\alpha}\right)\right)}
{\left(\left(\alpha\sinh\left(\frac{\tau}{2\alpha}\right)-i\epsilon\cosh\left(\frac{\tau}{2\alpha}\right)\right)^2-(na)^2\right)^2} \ .
\end{equation}
The contributions to the integral from each term in the sum
may be calculated separately by contour integration. The result is
\begin{eqnarray}
P(\omega) & = & 
\frac{1}{2\pi}
\frac{(\omega^2+\frac{1}{4\alpha^2})}{(1+e^{2\pi\alpha\omega})}
\nonumber
\\
&  &
+2
\sum_{n=1}^\infty
\frac{\eta^n}{(1+e^{2\pi\alpha\omega})}
\left[
\frac{\alpha^2n^2a^2\cos\left(2\omega\alpha\mathrm{arctanh}\left(\frac{(\alpha^2n^2a^2+n^4a^4)^{1/2}}{\alpha^2+n^2a^2}\right)\right)}
{4\pi\left(\frac{\alpha^2}{\alpha^2+n^2a^2}\right)^{1/2}(2n^6a^6+4n^4a^4\alpha^2+2n^2a^2\alpha^4)}
\nonumber \right.\\
&  & 
\label{eqn:accepowerm0}
\left.+\frac{\alpha^3(\alpha^2n^2a^2+n^4a^4)^{1/2}\omega\sin\left(2\omega\alpha\mathrm{arctanh}\left(\frac{(\alpha^2n^2a^2+n^4a^4)^{1/2}}{\alpha^2+n^2a^2}\right)\right)}
{2\pi\left(\frac{\alpha^2}{\alpha^2+n^2a^2}\right)^{1/2}(2n^6a^6+4n^4a^4\alpha^2+2n^2a^2\alpha^4)}
\right] \ ,
\end{eqnarray}
where again $\eta$ labels the spin structure. We see that the power spectrum
depends on the spin structure. The $n=0$ term in (\ref{eqn:accepowerm0}) agrees 
with the Minkowski space power spectrum (\ref{eqn:minkdiracpspec}) as expected, and both the
$n=0$ and $n>0$ terms in (\ref{eqn:accepowerm0}) contain the fermionic factor.
Note that no simple relation holds between the power spectrum (\ref{eqn:accepowerm0})
and the transition rate of the linearly coupled scalar field detector (\ref{M0response}),
in contrast to the relation we observed on the inertial worldline.

For the transition rate of a fermionic detector on $M_0$ we have
\begin{equation}
\dot{F}_\tau(\omega)=
2\int^{\infty}_{0}ds\,
Re\left(e^{-i\omega{s}}\mathrm{Tr}\left(S^+_{M_0}(\tau,\tau-s)^2\right)\right) \ .
\label{eqn:transm0dircexp}
\end{equation}
We may evaluate (\ref{eqn:transm0dircexp}) on the uniformly accelerated worldline
by substituting the worldline into (\ref{eq:diraccorM0}) and (\ref{eq:SeqG}).
It is easy to show that the $n=0$ term leads to the transition rate
found on Minkowski space (\ref{eq:diractranm}) as expected.
The evaluation of the other terms is not so straightforward as the
residues are not so easy to calculate. We shall not present the
result here.

\subsection{Dirac detector on $M_-$}
\label{sec-diracM-}

On $M_-$ we can again build expressions from those on 
$M$ (or $M_0$) via the method of images.
The transition rate is given by
\begin{equation}
\label{eq:transM-}
\dot{F}_\tau(\omega)=
2\int^{\infty}_{0}ds\,
Re\left(e^{-i\omega{s}}\mathrm{Tr}\left(S^+_{M_-}(\tau,\tau-s)^2\right)\right) \ ,
\end{equation}
and
\begin{eqnarray}
\label{eq:m-intermsm0}
S^+_{M_-}(\tau,\tau') & = & S^+_{M_0}(\tau,\tau')+\rho{S^+_{M_0}(\tau,J_-(\tau'))} \ ,
\\
                      & = & \sum_{n\in\Z}\rho^n{S^+_{M}(\tau,J_-^n(\tau'))} \ ,
\end{eqnarray}
where $S^+_{M_0}(\tau,\tau')$ and $S^+_{M}(\tau,\tau')$ are written in terms
of a vierbein which rotates by $2\pi$ in the $(x,y)$-plane as $z\mapsto{z+2a}$
(i.e. the one spin structure on $M_0$ compatible with the two on $M_-$
\cite{pl:langlois}). 
$\rho=1(-1)$ labels spinors with periodic (antiperiodic) boundary conditions
on $M_-$ with respect to this vierbein. That is $\rho$ labels the two possible 
spin structures on $M_-$.

Now on $M_-$ our main question of interest is whether or not
our detector can distinguish the 
two possible spin structures.
The stress tensor for the massless
spinor field in $M_-$ \cite{pl:langlois} has non-zero shear components, 
$\langle{0_-}|{T_{xz}}|0_-\rangle$ and $\langle{0_-}|{T_{yz}}|0_-\rangle$,
which change sign under a change of spin structure.
It is therefore conceivable that a detector with a non-zero $z$-component 
of angular momentum could detect the spin structure. However as the
relation between $\langle{0_-}|{T_{\mu\nu}}|0_-\rangle$ and the detector
response is not clear it is not possible to tell in advance whether or not our
detector model will be sensitive to the spin structure. 

We consider therefore a detector following the trajectory 
\begin{eqnarray}
t=t(\tau) \ ,\quad   x=x(\tau) \ ,\quad   y=y_0 \ ,\quad   z=z_0 \ ,
\label{eqn:m-trajdir}
\end{eqnarray}
where $y_0\neq{0}$ and $z_0$ are constants.
First we note that there is no direct 
analogue of the Rindler noise power spectrum here as the power spectrum is defined in 
\cite{tk:takagi} only for stationary trajectories. We therefore look
directly at the transition rate. 
From (\ref{eq:m-intermsm0}) and (\ref{eq:transM-})
the transition rate will contain four terms. 
The first term, coming from $\mathrm{Tr}\left(S^+_{M_0}(\tau,\tau')\right)^2$,
will give us the same response as on $M_0$ (for $\eta=-1$ in (\ref{eq:diraccorM0}),
as there expressions are written with respect to the standard Minkowski vierbein).
This part is independent of spin structure ($\rho$) on 
$M_-$. The fourth term, coming from $\mathrm{Tr}\left(\left(\rho{S^+_{M_0}}(\tau,J_-(\tau'))\right)^2\right)$,
will also be independent of spin structure, as it contains only
$\rho^2=1$ in both cases. 
Thus the only way in which the transition rate may be 
sensitive to the spin structure on $M_-$ is through the cross terms,
$\mathrm{Tr}(\rho{S^+_{M_0}(\tau,J_-(\tau'))}S^+_{M_0}(\tau,\tau'))$ and
$\mathrm{Tr}(\rho{S^+_{M_0}(\tau,\tau')}{S^+_{M_0}(\tau,J_-(\tau'))})$.
However it is a reasonably straightforward matter to show that these traces
are both $0$ on the trajectory (\ref{eqn:m-trajdir}), due to simple cancellations in the products of
the Wightman functions. Thus we see, even without an explicit calculation
on a specific trajectory, that the transition rate cannot depend on 
$\rho$, and so the detector is not sensitive to the spin structure,
for \emph{any} motion at constant $y$ and $z$.

Unfortunately an explict evaluation of the transition rate on the inertial
or uniformly accelerated worldlines, as on $M_0$, is difficult to obtain
and we shall not discuss it further here.


\section{Static detectors on the $\rp$ geon}
\label{sec:Geon}


In the recent literature Deser and 
Levin \cite{dl:deser3,dl:deser2,dl:deser}
have presented kinematical arguments for the calculation
of the Hawking-Unruh effects in a large class of black hole
and cosmological spaces by mapping the trajectories of 
detectors in these spacetimes to Rindler trajectories 
in higher dimensional embedding spaces (known as
GEMS, or global embedding Minkowski spacetimes) in which these
spacetimes have global embeddings. In \cite{dl:deser3} uniformly
accelerated observers in de Sitter and Anti de Sitter space are considered. It is
seen that in de Sitter space their experience is thermal with temperature $T=a_5/(2\pi)$ 
where $a_5$ is their associated acceleration in the $5$-dimensional embedding space.
In Anti de Sitter space their experience is thermal provided the acceleration
is above a certain threshold. In \cite{dl:deser2} static observers in
Schwarzschild space are considered via a $6$-dimensional flat embedding space 
and the expected temperature and entropy are recovered. In
\cite{dl:deser} this GEMS approach for the derivations of temperature
and entropy is extended to 
Schwarzschild-(anti) de Sitter and Reissner-Nordstr\"om spaces in four dimensions and rotating BTZ spaces
in three dimensions, and the methods of \cite{dl:deser} can
be readily adapted to other cases. We note that indeed any
Einstein geometry has a GEMS \cite{go:gems}.

\cite{sl:gems2} considers  
GEMS calculations on a large class of higher dimensional black
holes, generalising the four-dimensional results of Deser and Levin
(and the results for the four-dimensional AdS hole and others
in  \cite{hkp:sohkimpark}).
In particular, $d$-dimensional Schwarzschild and Reissner-Nordstrom
in asymptotically flat, de Sitter and Anti de Sitter spaces are discussed.
The case of four-dimensional asymptotically locally anti-de Sitter is
particularly interesting as solutions with planar, cylindrical,
toroidal and hyperbolic horizon topology exist. The higher dimensional
versions of these non-spherical AdS black holes are also considered.
Their global embeddings in higher dimensional Minkowski spaces are found
and the associated temperatures and entropies obtained.
Other references on GEMS come from the group 
of Hong, Park, Kim, Soh and Oh 
\cite{hkp:sohkimpark,hkp:sohkimpark1,hkp:sohkimpark1.5,hkp:sohkimpark2,hkp:sohkimpark3}. These include
the $4$-dimensional AdS hole as mentioned above, static rotating and charged BTZ holes,
$(2+1)$ de Sitter holes, scalar tensor theories, charged dilatonic
black holes in $1+1$ dimensions, charged and uncharged black
strings in (2+1) dimensions, and a few other cases.
A recent paper by Chen and Tian \cite{ct:nonstatgems} argues that
the GEMS approach holds for general stationary motions in curved spacetimes.
However these authors argue further that the approach in general fails for 
non-stationary motions. The example they use is that of a freely falling
observer in the Schwarzschild geometry. We note here that although their
argument does prove that the GEMS argument is not valid for some
non-stationary trajectories by use of an example, it does not prove that
the GEMS approach is useless for all such trajectories.

Within the kinematical arguments employed in all the work reviewed above
the great simplification in working with these GEMS is that
we are mapping situations in curved spacetimes to corresponding
ones in a flat spacetime, where calculations are always simpler,
both conceptually and technically.
It seems reasonable following the success of the GEMS programme
that the responses of particle detectors in black hole
and cosmological backgrounds
could also be related in some way to responses of 
corresponding detectors in their GEMS. We note 
immediately that such a mapping of detector responses is
clearly not trivial as we would expect different responses
to occur due to the different dimensions which the spacetimes and 
their GEMS have, however some relation is still expected.
In this section then we present an argument which should
be relevant to the response of a static detector in the 
single exterior of the $\rp$ geon black hole (and the Kruskal
spacetime) via an embedding of the Kruskal manifold into
a $7$-dimensional Minkowski space. This embedding space is 
different to the $6$-dimensional embedding of Kruskal
so far used in the GEMS literature \cite{fr:Fronsdal}, but we use
it as it is more easily adapted to the $\rp$ geon.

We begin by first presenting the embedding (see \cite{gb:embed}).
The complexified Kruskal manifold $M_C$ here is considered to be an algebraic variety 
in $\C^7$. With coordinates $(z_1,\ldots,z_7)$ and metric
\begin{equation}
ds^2
=
-(dz_1)^2-(dz_2)^2
-
\ldots
-(dz_6)^2
+(dz_7)^2 \ ,
\end{equation}
$z_7$ being the timelike coordinate, $M_C$ is determined by
\begin{eqnarray}
(z_6)^2-(z_7)^2+4/3(z_5)^2 & = & 16M^2 \ ,
\nonumber
\\
\left((z_1)^2+(z_2)^2+(z_3)^2\right)(z_5)^4 & = & 576M^6 \ ,
\nonumber
\\
\sqrt{3}z_4z_5+(z_5)^2 & = & 24M^2 \ .
\end{eqnarray}
The Lorentzian section of $M_C$, denoted by $\hat{M}_L$, is the subset
stabilised by $J_L:(z_1,\ldots,z_7)\mapsto(z_1^*,\ldots,z_7^*)$, where
$*$ stands for complex conjugation.
$\hat{M}_L$ consists of two connected components, one with $z_5>0$
and one $z_5<0$, both of which are isometric to the Kruskal manifold,
which we denote by $M_L$.
An explicit embedding of $M_L$ into $\hat{M}_L$ with $z_5>0$ is
given by
\begin{eqnarray}
z_1 & = & r\sin\theta\cos\phi \ ,
\nonumber
\\
z_2 & = & r\sin\theta\sin\phi \ ,
\nonumber
\\
z_3 & = & r\cos\theta \ ,
\nonumber
\\
z_4 & = & 4M\left(\frac{r}{2M}\right)^{1/2}-2M\left(\frac{2M}{r}\right)^{1/2} \ ,
\nonumber
\\
z_5 & = & 2M\left(\frac{6M}{r}\right)^{1/2} \ ,
\nonumber
\\
z_6 & = & 4M\left(\frac{2M}{r}\right)^{1/2}\mathrm{exp}\left(-\frac{r}{4M}\right)X \ ,
\nonumber
\\ 
z_7 & = & 4M\left(\frac{2M}{r}\right)^{1/2}\mathrm{exp}\left(-\frac{r}{4M}\right)T \ ,
\label{eqn:z1toz7}
\end{eqnarray}
with $X^2-T^2>-1$ and $r=r(T,X)$ defined as the unique solution to
\begin{equation}
\left(\frac{r}{2M}-1\right)\mathrm{exp}\left(\frac{r}{2M}\right)=X^2-T^2 \ .
\end{equation}
Here $(T,X,\theta,\phi)$ are a set of usual Kruskal 
coordinates, giving the usual Kruskal metric on 
$M_L$.
In each of the four regions of $M_L$, $|X|\neq|T|$, one
can introduce as usual local Schwarzschild coordinates
$(t,r,\theta,\phi)$. For $X>|T|$, the transformation reads
\begin{eqnarray}
T & = & \left(\frac{r}{2M}-1\right)^{1/2}\mathrm{exp}\left(\frac{r}{4M}\right)\sinh\left(\frac{t}{4M}\right) \ ,
\nonumber
\\
X & = & \left(\frac{r}{2M}-1\right)^{1/2}\mathrm{exp}\left(\frac{r}{4M}\right)\cosh\left(\frac{t}{4M}\right) \ ,
\end{eqnarray}
where $r>2M$, and the expressions for $z_1,z_2,\ldots,z_5$ are as in (\ref{eqn:z1toz7}) while those for $z_6$ and $z_7$ become
\begin{eqnarray}
z_6 & = & 4M\left(1-\frac{2M}{r}\right)^{1/2}\cosh\left(\frac{t}{4M}\right) \ ,
\nonumber
\\
z_7 & = & 4M\left(1-\frac{2M}{r}\right)^{1/2}\sinh\left(\frac{t}{4M}\right) \ .
\end{eqnarray}
Recalling that $z_7$ is the timelike coordinate in the embedding space, we
see immediately that an observer static in the exterior region $X>|T|$
at $r=\textrm{const}, \theta=\textrm{const}, \phi=\textrm{const}$ is a Rindler
observer in the $7$-dimensional embedding space with $(z_1,\ldots,z_5)$ constant
and acceleration in the $z_6$-direction of magnitude 
\begin{equation}
\label{eqn:accelembed}
a=1/\alpha=\frac{1}{4M\left(1-\frac{2M}{r}\right)^{1/2}} \ .
\end{equation}
As we have seen, the response of such a Rindler
detector in the embedding space is thermal with the
associated temperature
\begin{equation}
\label{eqn:localgemshawk}
T=\frac{a}{2\pi}=\frac{1}{2\pi\alpha}=\frac{1}{8\pi{M}\left(1-\frac{2M}{r}\right)^{1/2}} \ .
\end{equation}
This gives the Hawking temperature as seen by the static observer in the
black hole spacetime.
The associated black hole temperature, i.e. the temperature as seen at infinity,
in the Kruskal spacetime is
given by the Tolman relation
\begin{equation}
\label{eqn:infgems hawk}
T_0=g_{00}^{1/2}T=\frac{1}{8\pi{M}} \ ,
\end{equation}
(\ref{eqn:localgemshawk}) and (\ref{eqn:infgems hawk}) are the expected expressions
on Kruskal space \cite{bd:book}. Thus the black hole temperature as seen by a
static observer has been derived from the Unruh temperature seen by the
associated Rindler observer in the global embedding Minkowski spacetime.

Next we consider the $\rp$ geon. This is built as a quotient
of the Kruskal manifold under the involutive isometry
$J_G:(T,X,\theta,\phi)\mapsto(T,-X,\pi-\theta,\phi+\pi)$.
We now extend the action of the group generated by $J_G$ 
to the $7$-dimensional embedding space $M_C$ in order to
obtain a suitable embedding space for the geon.
A suitable extension of $J_G$ is
$\bar{J}_G:(z_1,z_2,z_3,z_4,z_5,z_6,z_7)\mapsto(-z_1,-z_2,-z_3,z_4,z_5,-z_6,z_7)$,
which is an involution on $M_C$.
Again the worldline of a static detector in the $\rp$ geon exterior $X>|T|$ is
mapped to the worldline of a Rindler observer with acceleration in the
$z_6$-direction with magnitude (\ref{eqn:accelembed}) in this embedding space.
We suggest therefore that the calculations of the time dependent responses
of an accelerated observer in the $d$-dimensional quotients of Minkowski space, 
done in section \ref{sec:quotient}, should have relevance to the response of a static detector in the
exterior of the $\rp$ geon (although the exact nature of the relationship is 
not clear).
In particular if we specialise the results of section \ref{sec:quotient} to 
a detector with uniform acceleration (\ref{eqn:accelembed}) in the 
quotient of a $7$-dimensional Minkowski space under involution $\bar{J}_G$ we
see that the response has two parts. The thermal time-independent part
is given by (\ref{thermal}), which in the present case reads
\begin{equation}
\dot{F}_{M\tau}(\omega)
=
\frac{a^4}{64\pi^2(e^{\frac{2\pi\omega}{a}}+1)}
(1/4+\omega^2/a^2)(9/4+\omega^2/a^2) \ .
\end{equation}
Again this is a thermal response associated with a temperature
$T=\frac{a}{2\pi}$. Clearly the response is different to that of the
static detector on 
Kruskal due to the higher number of dimensions though the two should be related.
The image part of the response for this Rindler detector is, 
in the case of a detector switched on in the infinite past,
\begin{equation}
\dot{F}_{I\tau}(\omega)
=
\frac
{3}
{8\pi^{3}}
\int_0^\infty
ds
\;
\frac{\cos(\omega{s})}
{\left(\frac{4}{a^2}\cosh^2\left(\frac{a(2\tau-s)}{2}\right)+4r^2\right)^{5/2}} \ .
\end{equation}

\begin{figure}[htbp]
\includegraphics[angle=0, width={4in}]{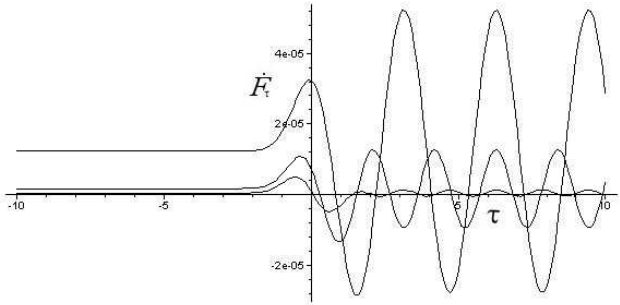}
\centering
\caption[Transition rate associated with the response of a static detector on the $\rp$ geon.]{Transition rate 
for a detector uniformly accelerated in the $z_1$ direction on the
quotient of $7$-dimensional Minkowski space under the involution $\bar{J}_G$.
The parameters are $\alpha=1$, $C=(2z_1)^2+(2z_2)^2+(2z_3)^2=4r^2=64/9$, and $\omega=1$ (upper curve), 
$\omega=1.5$ and $\omega=2$ (lower curve).}
\label{fig:GEMSGEON}
\end{figure}

The total response for certain values of the parameters is shown in figure \ref{fig:GEMSGEON}.
The comments of section \ref{sec:autoacelmbound} then follow. The image part
consists of a term periodic in $\tau$ with period $\pi/\omega$ plus a term
bounded by a function which dies off exponentially for large $\tau$.
The numerical evidence exhibits behaviour
qualitatively very similar to that of figure \ref{fig:accel}.
The comments made in the section \ref{sec:autoacelmbound} about finite time 
detections also follow here. In particular the oscillatory behaviour of the 
boundary part as $\tau\rightarrow\infty$ is a property only of the case of infinite time detection.
For a detector switched on at $-\infty<\tau_0<0$ the transition rate oscillates for some
time period with $\tau>0$, but eventually it will fall to the thermal response and so
at late times the difference between the response on the Minkowski space
and that on the quotient space vanishes. 
That is, for instantaneous, exponential and Gaussian switching the image term behaves qualitatively 
as shown in figures \ref{fig:finitetime}, \ref{fig:finitetimewindow} and \ref{fig:finitetimewindow2}, respectively. 
This implies for finite time static detectors 
on the $\rp$ geon the difference between the response there and that on Kruskal spacetime
falls off also to $0$ at late times, which is in agreement with the comments made in \cite{lm:geon}.
Note however the different behaviour in the case of the infinite time detection.

\section{De Sitter and $\rp$ de Sitter spaces}
\label{sec-desitrp3}

In this section we begin in subsection \ref{sec-desit} by considering a model detector in de Sitter space. 
Throughout we consider a conformally coupled massless scalar field moving 
through the Euclidean vacuum \cite{ba:desitstates}.
We see that here a similar situation is
encountered to that found by Schlicht for the uniformly accelerated
detector in Minkowski space.
It is seen that in the case of a comoving detector switched on in
the infinite past and off at finite $\tau$, if the
correlation function is regularized by a naive $i\epsilon$-prescription, as for
example is done in Birrell and Davies~\cite{bd:book}, we are led to an unphysical,
$\tau$-dependent response. 
We therefore introduce an alternative regularization. 
Further, for a comoving detector,
we show that such a regularization can arise also by considering a model 
detector with spatial extent, that is by considering
a smeared field operator in the interaction Hamiltonian.
We recover the usual time independent thermal response for comoving 
and uniformly accelerated detectors.

Subsection \ref{sec:rp3} then considers comoving observers in $\rp$ de Sitter space,
such that the motion is orthogonal to the distinguished foliation \cite{jk:desit}.
In addition to the thermal part seen in de Sitter space, the transition rate contains
an image part, related to that found in section \ref{sec:minkbound} for a uniformly accelerated 
detector on a four-dimensional Minkowski space with a planar boundary.
We also address a comoving detector in de Sitter and $\rp$ de Sitter space in a GEMS approach, by considering
the response of the associated uniformly accelerated detectors in higher dimensional
Minkowski (with boundaries in the case of $\rp$ de Sitter) embedding spaces.
As we are able to do the calculations both in the original curved spaces
and in the global embedding spaces, the results help to clarify the
relation and validity of relating detector responses to those in embedding spaces.

\subsection{Detectors in de Sitter space}
\label{sec-desit}

We represent $d$-dimensional de Sitter space as the hyperboloid 
\begin{equation}
\label{eqn:desithyper}
z_0^2-z_1^2-\cdots-z_{d}^2=-\alpha^2 \ ,
\end{equation}
embedded in the ${d+1}$-dimensional Minkowski space,
\begin{equation}
ds^2=
dz_0^2-dz_1^2-\cdots-dz_{d}^2 \ ,
\end{equation}
with $z_i$ real-valued coordinates.
Let us consider the coordinates $(t,\mathbf{x})$ defined by
\begin{eqnarray}
z_0 & = & \alpha\sinh(t/\alpha)+\frac{e^{t/\alpha}}{2\alpha}|\mathbf{x}|^2 \ ,
\nonumber
\\
z_{d} & = & \alpha\cosh(t/\alpha)-\frac{e^{t/\alpha}}{2\alpha}|\mathbf{x}|^2 \ ,
\nonumber
\\
z_i & = & e^{t/\alpha}x_i \ .
\end{eqnarray}
These coordinates cover the half of the de Sitter hyperboloid
given by $z_0+z_{d}>0$.
The line element is that of a $d$-dimensional
Friedman-Robertson-Walker spacetime with exponentially expanding flat
spatial sections,
\begin{equation}
ds^2
=
dt^2
-
e^{2t/\alpha}
\left(dx_1^2+\cdots+d_{d-1}^2\right) \ .
\end{equation}
Introducing the conformal time $\eta=-\alpha{e^{-t/\alpha}}$, the line
element becomes conformal to Minkowski space,
\begin{equation}
\label{eqn:conflineel}
ds^2
=
\frac{\alpha^2}{\eta^2}\left[d\eta^2-\sum_{i=1}^{d-1}(dx_i)^2\right] \ ,
\end{equation}
where $-\infty<\eta<0$.

Consider a massless conformally coupled scalar field in the
line element (\ref{eqn:conflineel}). 
A complete set of mode solutions to the Klein-Gordon equation positive 
frequency with respect to conformal Killing time $\eta$ is given by
\cite{bd:book}
\begin{equation}
\label{eqn:conformspacemodesol}
\phi_\mathbf{k}(\eta,\mathbf{x})
=
\frac{\eta^{d/2-1}}{(2\alpha^{d-2}\omega(2\pi)^{d-1})^{1/2}}
e^{-i\omega\eta+i\mathbf{k}\cdot\mathbf{x}} \ .
\end{equation}
The field may be expanded in the modes (\ref{eqn:conformspacemodesol}) and quantized
in the usual way.
The associated vacuum state, that is the conformal
vacuum, coincides with the state known as the Euclidean vacuum~\cite{bd:book}.
The Euclidean vacuum $|0_E\rangle$ is uniquely characterised
as the state whose correlation function $\langle{0_E}|\phi(x)\phi(x')|0_E\rangle$
is invariant under the connected component of the de Sitter group,
and the only singularities of the correlation function are when $x'$ is
on the lightcone of $x$~\cite{ba:desitstates}.
Even though we have here defined the Euclidean vacuum in
coordinates that only cover half of the de Sitter hyperboloid, it 
is worth mentioning that the state is well defined on the whole hyperboloid~\cite{bd:book}.

We now consider a monopole detector linearly coupled to the field 
via the interaction Hamiltonian
\begin{equation}
H_{\mathrm{int}}=cm(\tau)\phi(x(\tau)) \ .
\end{equation}
The transition rate, for a detector originally in state $|E_0\rangle$ with the field
in the Euclidean vacuum state at time $\tau_0$, to be found in the state $|E_1\rangle$
at time $\tau>\tau_0$ is then, to first order in perturbation theory
\begin{equation}
\dot{F}_{\tau}(\omega)=
2\int^{\tau-\tau_0}_{0}ds\,
Re
\left(
e^{-i\omega{s}}\langle{0_E}|\phi(\tau)\phi(\tau-s)|0_E\rangle
\right) \ ,
\label{eqn:conftransrate}
\end{equation}
where $\omega=E_1-E_0$.
The correlation function $\langle{0_E}|\phi(x)\phi(x')|0_E\rangle$ from (\ref{eqn:conformspacemodesol})
is given by
\begin{eqnarray}
\langle{0_E}|\phi(x(\tau))\phi(x(\tau'))|0_E\rangle
& = & 
\frac{(\eta\eta')^{d/2-1}}
{\alpha^{d-2}(2\pi)^{d-1}}
\nonumber
\\
\label{eqn:correconf}
&    &
\times
\int
\frac{d^{d-1}k}{2|\mathbf{k}|}\,
e^{-i|\mathbf{k}|(\eta(\tau)-\eta(\tau'))+i\mathbf{k}\cdot(\mathbf{x}(\tau)-\mathbf{x}(\tau'))} \ .
\end{eqnarray}
(\ref{eqn:correconf}) is conformally related to the Minkowski space Wightman
function by
\begin{equation}
\label{eqn:conrelat}
\langle{0_E}|\phi(x)\phi(x')|0_E\rangle
=
\left(\frac{\eta}{\alpha}\right)^{d/2-1}
\langle{0}|{\phi}(x){\phi}(x')|0\rangle
\left(\frac{\eta'}{\alpha}\right)^{d/2-1} \ .
\end{equation}
The integrals in (\ref{eqn:correconf}) may be performed by
transforming to hyperspherical coordinates. The integral over $|\mathbf{k}|$
requires regularization.

We shall now specialize to four-dimensional de Sitter space (although the
extension to higher dimensions is straightforward). 
If we regularize (\ref{eqn:correconf}) with a naive $i\epsilon$-prescription, that is,
we introduce the cut-off $e^{-\epsilon\omega}$, we find via a numerical calculation that
the transition rate (\ref{eqn:conftransrate}) for a comoving detector with worldline
$t=\tau$, $\mathbf{x}=0$, when the detector is switched on at $\tau=-\infty$ and
off at $\tau$, is time dependent and therefore apparently unphysical.
We are led, as was Schlicht with the uniformly accelerated detector in Minkowski
space, to an alternative regularization of (\ref{eqn:correconf}). 
Our proposal is to consider the correlation function with the relation
(\ref{eqn:conrelat}) to the Minkowski space correlation function of Schlicht
\cite{sc:schlicht}.
That is 
\begin{eqnarray}
& \langle{0}_E|\phi(\tau)\phi(\tau')|0_E\rangle=
-\frac{1}{4\pi^2\alpha^2}
\frac{\eta(\tau)\eta'(\tau)}{A} \nonumber
\\
& A=[(\eta(\tau)-\eta(\tau')-i\epsilon(\dot{\eta}(\tau)+\dot{\eta}(\tau'))^2
-(\mathbf{x}(\tau)-\mathbf{x}(\tau')-i\epsilon(\dot{\mathbf{x}}(\tau)+\dot{\mathbf{x}}(\tau')))^2] \ ,
\nonumber
\\
\label{eq:euclidcorre}
\end{eqnarray}
with the transition rate still given by (\ref{eqn:conftransrate}).

Consider a uniformly accelerated detector following the worldline
\begin{eqnarray}
z_0 & = & \alpha\sinh(t/\alpha)+\frac{e^{-t/\alpha}}{2\alpha}r^2 \ ,
\nonumber
\\
z_4 & = & \alpha\cosh(t/\alpha)-\frac{e^{-t/\alpha}}{2\alpha}r^2 \ ,
\nonumber
\\
z_1 & = & z_2  = 0 \ ,
\nonumber
\\
\label{eqn:accelobs}
z_3 & = & r \ ,
\end{eqnarray}
with $r=\textrm{constant}$. The worldline of such an observer in the 
embedding space is a hyperbola $(z_4)^2-(z_0)^2=\alpha^2-r^2$, $z_1=z_2=0$, $z_3=r$.
In the de Sitter space the observer has constant proper
acceleration $a$, where $a^2=-g_{\mu\nu}\dot{u}^\mu\dot{u}^\nu$,
$\dot{u}^\mu={u}^\nu(\nabla_\nu{u}^\mu)$ and $u^\mu$ is the tangent vector of the
trajectory, of magnitude
\begin{equation}
a=\frac{r}{\alpha(\alpha^2-r^2)^{1/2}} \ .
\end{equation}
The proper time for the accelerated observer is
$\tau=(\alpha^2-r^2)^{1/2}t/\alpha$.

Substituting (\ref{eqn:accelobs}) into (\ref{eq:euclidcorre}) we find 
\begin{equation}
\langle{0}_E|\phi(\tau)\phi(\tau')|0_E\rangle=
-\frac{1}{16\pi^2\left((\alpha^2-r^2)^{1/2}\sinh\left(\frac{\tau-\tau'}{2(\alpha^2-r^2)^{1/2}}\right)
-i\epsilon\cosh\left(\frac{\tau-\tau'}{2(\alpha^2-r^2)^{1/2}}\right)\right)^2} \ ,
\end{equation}
and from (\ref{eqn:conftransrate}) the transition rate
of the detector switched on in the infinite past and off at time $\tau$ is 
independent of $\tau$ and is given by
\begin{equation}
\dot{F}_{\tau}(\omega)=
\frac{\omega}
{2\pi(e^{2\pi\omega(\alpha^2-r^2)^{1/2}}-1)} \ .
\end{equation}
The accelerated detector thus experiences a
thermal response at temperature
\begin{equation}
T=\frac{1}{2\pi}
\left(\frac{1}{\alpha^2}+a^2\right)^{1/2} \ .
\end{equation}
The response of a comoving detector is obtained by setting $a=0$. The transition rate
is still thermal at temperature
$T=1/(2\pi\alpha)$.
These results agree with the previous literature (e.g \cite{bd:book,dl:deser3}).
What is new is that we have obtained these results in a causal way for a 
detector switched on in the infinite past and read at a finite time, as opposed to the
case usually considered of a detection over the entire worldline.

We end this section by showing that the regularization in (\ref{eq:euclidcorre}), in
the case of a comoving observer, may be obtained by considering the monopole detector as the
limit of an extended 
detector in de Sitter space.
The reason why this is simple in the case of a comoving observer but not for
other trajectories is that spatial hypersurfaces of constant $t$ in the
coordinates $(t,\mathbf{x})$ are flat Euclidean spaces, which allow us to 
introduce a detector with infinite spatial extent along these slices. 
Care must be taken when defining the shape function for the detector
however, because the hypersurfaces are expanding with increasing $t$,
with a shape which is rigid in the proper distance the regularization follows.
The averaging over spatial hypersurfaces in effect introduces a short distance,
high frequency cut-off in the modes. 

The detector model is that of section \ref{sec-linmink}. It is a multi-level
quantum mechanical system coupled to a massless conformally coupled
scalar field via the interaction Hamiltonian
\begin{equation}
\label{eqn:comovhint}
H_{\mathrm{int}}=cm(\tau)\phi(\tau) \ .
\end{equation}
We consider a detector following the trajectory
$t=\tau$ that is $\eta=-\alpha{e^{-\tau/\alpha}}$
and $\mathbf{x}=0$. In (\ref{eqn:comovhint})
we consider the field smeared with a detector profile
function over constant $\tau$ hypersurfaces, that is
\begin{equation}
\label{eqn:comovshape}
\phi(\tau)=\int
d^3x\,
W_\epsilon(\mathbf{x})
\phi(\tau,\mathbf{x}) \ .
\end{equation}
For the profile function we choose
\begin{equation}
\label{eq:shapedesitt}
W_{\epsilon}(\mathbf{x})
=
\frac{1}{\pi^2}
\frac{\epsilon{e^{-\tau/\alpha}}}{(\mathbf{x}^2+\epsilon^2{e^{-2\tau/\alpha}})^{2}} \ .
\end{equation}
The detector shape (\ref{eq:shapedesitt}) is now time dependent! The reason for this
is that we want a detector which is rigid in its rest frame.
That is, we want a detector which is rigid with
respect to proper distance and not comoving distance.
The two distances are related by
$L_{\textrm{prop}}=e^{\tau/\alpha}L_{\textrm{comov}}$.
In (\ref{eqn:comovshape}) the integration is done over
$\mathbf{x}$, which is a comoving coordinate, and using 
a time independent shape function there would mean that
the detector is rigid with respect to comoving distance.
It is a simple matter to show that a shape function which selects
a distance scale $L'$ may be obtained from one which selects
a distance scale $L$ by
\begin{equation}
W_{L'}(\mathbf{x})=\frac{L^3}{{L'}^3}
W_L\left(\frac{L}{L'}\mathbf{x}\right) \ .
\end{equation}
If we write (\ref{eq:shapedesitt}) now in terms
of proper distance we find
\begin{equation}
W_{\epsilon_{\textrm{prop}}}(\mathbf{\xi})
=
\frac{1}{\pi^2}
\frac{\epsilon_{\textrm{prop}}}{\left(\mathbf{x}^2+\epsilon^2_{\textrm{prop}}\right)^{2}} \ ,
\end{equation}
and so in terms of proper distance (\ref{eq:shapedesitt}) is
a rigid shape in the sense that it is time independent.
Using this shape function we find, using the mode expansion
of $\phi$,
\begin{eqnarray}
\langle{0}_E|\phi(\tau)\phi(\tau')|0_E\rangle
& = &
\frac{\eta\eta'}{(2\pi)^{3}\alpha^2}
\int
\frac{d^{3}k}{2\omega}
\int
d^{3}x\,
W_\epsilon(\mathbf{x})
e^{-i(\omega{\eta(\tau)}-\mathbf{k}\cdot\mathbf{x})}
\nonumber
\\
& & \times\int
d^{3}x'\,
W_\epsilon(\mathbf{x'})
e^{i(\omega{\eta(\tau')}-\mathbf{k}\cdot\mathbf{x})} \ .
\end{eqnarray}
Further we find
\begin{eqnarray}
\int
d^{3}x\,
W_\epsilon(\mathbf{x})
e^{-i(\omega{\eta(\tau)}-\mathbf{k}\cdot\mathbf{x})} 
& = & e^{-i|\mathbf{k}|\eta(\tau)}e^{-\epsilon|\mathbf{k}|e^{-\tau/\alpha}} \ ,
\nonumber
\\
& = & e^{-i|\mathbf{k}|\eta(\tau)}e^{-\epsilon|\mathbf{k}|\dot{\eta}(\tau)}
 \ ,
\end{eqnarray}
where the integration is done by transforming to spherical coordinates.
Hence
\begin{equation}
\label{eqn:corrdesitexten}
\langle{0}_E|\phi(\tau)\phi(\tau')|0_E\rangle
 = 
\frac{\eta\eta'}{(2\pi)^{3}\alpha^2}
\int
\frac{d^{3}k}{2\omega}\,
e^{-i\omega(\eta-\eta'-i\epsilon(\dot{\eta}+\dot{\eta}'))} \ .
\end{equation}
The expression (\ref{eqn:corrdesitexten}) agrees with that found above from
the ultraviolet cut-off regularization.

\subsection{$\rp$ de Sitter space}
\label{sec:rp3}

In this section we consider an inertial detector
that is linearly coupled to a conformally coupled massless scalar field in
$\rp$ de Sitter spacetime~\cite{jk:desit}.~\footnote{See also \cite{bm:mcinnes}
for a nice discussion on de Sitter space vs $\rp$ de Sitter.}

$\rp$ de Sitter space is built as a quotient of de Sitter space
under the group generated by the discrete isometry
\begin{equation}
\label{eqn:rpgroup}
J:(z_0,z_1,z_2,z_3,z_4)\mapsto(z_0,-z_1,-z_2,-z_3,-z_4) \ ,
\end{equation}
which induces a map $\tilde{J}$ on the hyperboloid (\ref{eqn:desithyper}).
Although $J$ has fixed points on $M$, $\tilde{J}$ acts freely on the
hyperboloid.
The isometry group of four-dimensional de Sitter space is $O(1,4)$, being the largest subgroup 
of the isometry group of the five-dimensional Minkowski embedding space which preserves (\ref{eqn:desithyper}).
The isometry group of $\rp$ de Sitter space is then the largest subgroup of $O(1,4)$ which
commutes with $J$. That is, $\Z_2\times{O(4)}/\Z_2$ where the non-trivial element
of the first $\Z_2$ factor sends $z_0$ to $-z_0$ while the non-trivial element
of the $\Z_2$ in the second factor is given by $J$ which clearly acts trivially on
$\rp$ de Sitter space. The connected component of the isometry group is 
$SO(4)$. The foliation given by $z_0=\mathrm{constant}$ hypersurfaces
is a geometrically distinguished one as it is the only foliation whose
spacelike hypersurfaces are orbits of the connected component of
the isometry group. This is made clearer by introducing the
globally defined coordinates $(t,\chi,\theta,\phi)$ 
\begin{eqnarray}
z_0 & = & \alpha\sinh(t/\alpha) \ ,
\nonumber
\\
z_{4} & = & \alpha\cosh(t/\alpha)\cos\chi \ ,
\nonumber
\\
z_1 & = & \alpha\cosh(t/\alpha)\sin\chi\cos\theta \ ,
\nonumber
\\
z_2 & = & \alpha\cosh(t/\alpha)\sin\chi\sin\theta\cos\phi \ ,
\nonumber
\\
z_3 & = & \alpha\cosh(t/\alpha)\sin\chi\sin\theta\sin\phi \ ,
\end{eqnarray}
in which the metric reads
\begin{equation}
ds^2
=
dt^2
-
\alpha^2\cosh^2(t/\alpha)
[d\chi^2+\sin^2\chi(d\theta^2+\sin^2\theta\,{d\phi^2})] \ ,
\end{equation} 
where $(\chi,\theta,\phi)$ on de Sitter ($\rp$ de Sitter) space are
a set of hyperspherical coordinates on $S^3$ ($\rp$) respectively.
$(t,\chi,\theta,\phi)$ make manifest the $O(4)$ isometry subgroup.

We denote by $|0_G\rangle$ the vacuum state induced by the
Euclidean vacuum $|0_E\rangle$ on de Sitter space (see \cite{jk:desit} for more details).
We consider a particle detector that is linearly coupled to a massless
conformally coupled scalar field. 
The detector and field are assumed to be in the states $|E_0\rangle$ and 
$|0_G\rangle$ respectively at time $\tau_0=-\infty$, and we seek the 
probability that at time $\tau>\tau_0$ the detector is found in the
state $|E_1\rangle$. Through arguments analogous to those in section \ref{sec-linmink},
the transition rate is 
\begin{equation}
\label{eqn:desittrans}
\dot{F}_{\tau}(\omega)=
2\int^{\infty}_{0}ds\,
Re
\left(
e^{-i\omega{s}}\langle{0_G}|\phi(\tau)\phi(\tau-s)|0_G\rangle
\right) \ .
\end{equation}
By the method of images we have
\begin{equation}
\label{eqn:methoddesit}
\langle{0_G}|\phi(x)\phi(x')|0_G\rangle
=
\langle{0_E}|\phi(x)\phi(x')|0_E\rangle
+
\langle{0_E}|\phi(x)\phi(Jx')|0_E\rangle \ ,
\end{equation}
where on the RHS expressions live in de Sitter space, and the correlation function
$\langle{0_E}|\phi(x)\phi(x')|0_E\rangle$ is given by  (\ref{eq:euclidcorre}).

Consider now a detector that follows the geodesic worldline 
\begin{eqnarray}
z_0 & = & \alpha\sinh(\tau/\alpha) \ ,\nonumber \\
z_4 & = & \alpha\cosh(\tau/\alpha) \ , \nonumber \\
\label{eqn:geodtraj}z_1 & = z_2 = z_3 =0 \ .
\end{eqnarray}
On $\rp$ de Sitter space this represents the motion of any geodesic observer
whose motion is orthogonal to the distinguished foliation.
The transition rate (\ref{eqn:desittrans}) is in two parts, a de Sitter part
and an image part. We have calculated already the de Sitter part, coming from
the first term in (\ref{eqn:methoddesit}), in 
section \ref{sec-desit}. The result was the usual thermal, Planckian, response at 
temperature $T=1/(2\pi\alpha)$.
We need the image term. In order to find $\langle{0_E}|\phi(x)\phi(Jx')|0_E\rangle$
on this worldline we first write $\langle{0_E}|\phi(x)\phi(x')|0_E\rangle$ in terms of the
coordinates $(z_0,z_1,z_2,z_3,z_4)$ of the embedding space and then act on $x'$ with $J$, finding
\begin{equation}
\label{eqn:desitcorr}
\langle{0_E}|\phi(x)\phi(Jx')|0_E\rangle
=\frac{1}{8\pi^2}
\frac{1}{\left(
+\mathbf{z}\cdot\mathbf{z'}
+z_0z'_0+\alpha^2\right)} \ ,
\end{equation}
where $\mathbf{z}=(z_1,z_2,z_3,z_4)$. The regularization has been omitted
as the wordline and its image under $J$ are completely 
spacelike separated.
The image term gives to the transition rate the contribution
\begin{equation}
\label{eqn:imagerpdesit}
\dot{F}_{I\tau}(\omega)=
2\int^{\infty}_{0}ds\,
\frac{\cos(\omega{s})}
{16\pi^2\alpha^2\cosh^2\left(\frac{2\tau-s}{2\alpha}\right)} \ .
\end{equation}

We see that the image term contribution (\ref{eqn:imagerpdesit}) is
exactly the same as the image term contribution in the response of a uniformly 
accelerated detector on a four-dimensional Minkowski space with boundary 
at $x=0$ ((\ref{eqn:boundaryaccel})
with $d=4$).
Therefore the total response of our inertial detector 
in $\rp$ de Sitter space (with Ricci scalar $R=12/\alpha^2$) is identical to the response
of a uniformly accelerated detector travelling in four-dimensional Minkowski space with
a boundary at $x=0$ with the acceleration $1/\alpha$ perpendicular
to the boundary.
It follows that our numerical results in figure \ref{fig:accel} 
also give the response on $\rp$ de Sitter space, with the 
appropriate interpretation for $\alpha$. In particular, 
the image term breaks the KMS condition and the 
response is non-thermal and non-Planckian. When the detector is switched
on in the infinite past, the response
at large $\tau$ is oscillatory
in $\tau$ with period $\pi/\omega$. 
When the detector is switched on at a finite time $\tau_0$, 
$\tau_0\ll{-1}$, numerical evidence indicates that
the response is approximately periodic in the region
$0<\tau<-\tau_0$, with period $\pi/\omega$, but it falls 
to the thermal response as $\tau\rightarrow\infty$,
as discussed further in section \ref{sec:minkbound}
and illustrared in figures \ref{fig:finitetime}, \ref{fig:finitetimewindow} and \ref{fig:finitetimewindow2}. 
This clarifies
and adds to the discussion given in~\cite{jk:desit}.

\begin{figure}[htbp]
\includegraphics[angle=0, width={4in}]{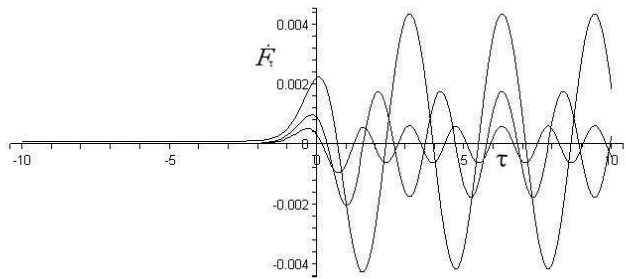}
\centering
\caption[Transition rate associated with the response of a comoving detector on $\rp$ de Sitter space.]{Transition rate 
for a detector uniformly accelerated in the $z_1$ direction
on the quotient space of $5$-dimensional Minkowski space under the involution $J$.
The parameters are $\alpha=1$, $C=(2z_1)^2+(2z_2)^2+(2z_3)^2=0$ and $\omega=1$ (upper curve), 
$\omega=1.5$ and $\omega=2$ (lower curve).}
\label{fig:GEMSDESITTER}
\end{figure}

We wish to compare these particle detector results in 
$\rp$ de Sitter space to the asociated GEMS particle detector.
We see from (\ref{eqn:geodtraj}) that the GEMS worldline of interest 
is a Rindler trajectory with acceleration $a=1/\alpha$ in the $5$-dimensional
embedding space. Therefore in the $5$-dimensional Minkowski embedding space of de Sitter space
we see that the response of
a detector following this worldline is a thermal one 
with associated temperature $T=1/(2\pi\alpha)$ and so we expect, as indeed we saw in
section \ref{sec-desit},
the response of the detector in de Sitter space to also be a thermal one with
temperature $T=1/(2\pi\alpha)$.
Again the actual responses of detectors in the two situations are not
identical (as seen in sections \ref{sec-linmink} and \ref{sec-desit}) due 
to the different dimensions of the spaces, the most obvious difference being the 
presence of the Planckian factor in the de Sitter response and the
Fermi factor in Rindler response on the odd dimensional embedding space. 
As the $\rp$ de Sitter spacetime is built as a quotient of
de Sitter space under the map $J:(z_0,z_1,z_2,z_3,z_4)\mapsto(z_0,-z_1,-z_2,-z_3,-z_4)$
we have immediately the action of this map on the embedding space.
The geodesic worldline of interest maps to a Rindler worldline in this embedding space
with acceleration $a=1/\alpha$,
so in the GEMS approach we consider a Rindler particle detector with this acceleration
in this $5$-dimensional embedding space. The transition rate 
was found in section \ref{sec:quotient}. The thermal part of the transition rate 
is constant in time and is given by
\begin{equation}
\dot{F}_{M\tau}(\omega)
=
\frac{a^2}{8\pi(e^{\frac{2\pi\omega}{a}}+1)}
(1/4+\omega^2/a^2) \ ,
\end{equation}
corresponding to the temperature
$T=\frac{a}{2\pi}$.
The image part of the transition rate depends on the proper time and is given by
\begin{equation}
\dot{F}_{I\tau}(\omega)
=
\frac
{1}
{4\pi^{2}}
\int_0^\infty
ds
\,
\frac{\cos(\omega{s})}
{\left(4/a^2\cosh^2\left(\frac{a(2\tau-s)}{2}\right)\right)^{3/2}} \ .
\end{equation}
The total response is shown for various values of the parameters
in figure \ref{fig:GEMSDESITTER}.
The qualitative similarities to the $\rp$ de Sitter transition rate are apparent.
They provide evidence that, at least in some cases, the GEMS procedure
may be applied to quotient spaces such as $\rp$ de Sitter space and
the $\rp$ geon where the embedding spaces are Minkowski spaces with suitable
identifications.

\clearpage

\section{Discussion}

In this paper we considered particle detector models in the
context of quantum field theory in curved spacetime. In particular we 
investigated the model of Schlicht \cite{sc:schlicht}. The 
model is that of a monopole detector linearly coupled to a massless scalar
field which is smeared with a window function in order to regularize the
correlation function in the transition rate.
We extended the regularization of the correlation function for the massless
linearly coupled scalar field to $d$-dimensional Minkowski space, and we showed
that it leads to the expected responses for inertial and uniformly accelerated
detectors switched on in the infinite past and off at $\tau<\infty$. Further we
extended the regularization of Schlicht to the massive scalar field in 
Minkowski space and have shown that it reduces to that of \cite{sc:schlicht}
in the massless limit. 

Next we introduced a model of a linearly coupled
massless scalar field detector on
spacetimes built as quotients of Minkowski space under certain discrete
isometries. In a number of cases the model, when switched on at $\tau_0=-\infty$
and read at $\tau<\infty$, was shown to reproduce the known asymptotic
responses. These cases include the uniformly accelerated detector on $M_0$ 
\cite{lm:geon,dlo:detecbound} as
well as the inertial and uniformly accelerated detectors on Minkowski space
with boundary when the motion is parallel to the boundary \cite{dlo:detecbound}.
These results suggest that our model is reasonable.
Further we presented a number of new responses, the most interesting of
which are the time dependent responses on Minkowski space with boundary and
on $M_-$. An inertial detector approaching the boundary on Minkowski space with
boundary was seen to react in a qualitatively similar way to one
travelling parallel to the boundary but taking progressively smaller distances
(ie comparing figures \ref{fig:inertialparal} and \ref{fig:inertialwl0}).
The main difference is that in the detector approaching the boundary a divergence
in the transition rate occurs as the boundary is reached.
A detector with uniform acceleration perpendicular to the boundary was
also considered, the results were seen to be more subtle and an interesting observation made. For
a detector which is switched on in the infinite past the transition rate 
is found to oscillate in $\tau$ at late times with period $\pi/\omega$, 
never tending to the Minkowski
thermal response, no matter how far from the boundary the detector
gets in the future. However for a detector switched on at a finite time
(that is $\tau_0>-\infty$), the response will at late times tend to the
thermal Minkowski response. For instantaneous, exponential and Gaussian 
switching functions the conclusion is the same.

Responses were also considered on the quotient spaces of Minkowski
space under the involution 
$J_{c_k}:(t,x_1,x_2,\ldots\,x_{d-1})\mapsto(t,-x_1,-x_2,-\ldots\,-x_k,x_{k+1},\ldots,x_{d-1})$
where $1<k<d$ and certain relations to the responses on Minkowski space with 
boundary noted.
The responses are relevent for discussions of detectors in the spacetime
outside an infinitely long and zero radius cosmic string.
The responses of uniformly accelerated detectors, where the motion is
in the $x_1$ direction, are also relevant for the discussion of static detectors
in the $\rp$ geon exterior as well as comoving detectors in $\rp$ de Sitter
via their global embedding Minkowski spacetimes (GEMS).

Next we extended the detector model and regularization to the massless Dirac field.
With a few minor technicalities the extension is quite 
straightforward.
The transition
rate and the power spectrum of the Dirac noise for a detector switched 
on in the infinite past on inertial and a uniformly accelerated trajectories was obtained.  
The power spectrum for the accelerated detector
agrees with the previous literature (see e.g \cite{tk:takagi}) 
and so suggests our model is reasonable. Further we briefly considered
the response of the Dirac detector on $M_0$ and $M_-$.
One aim was to see whether a uniformly accelerated Dirac detector
on $M_-$ could distinguish the two spin structures there. We 
found this not to be the case for our detector model.

In section \ref{sec:Geon} we considered the response of a static detector
in the exterior region of the $\rp$ geon via a global embedding Minkowski
space. Although the GEMS programme has so far only been applied in
a kinematical setting, our aim was to examine the possibility that the response of the detector
in the embedding space is related to that in the underlying curved space and further
whether the GEMS scheme can be applied to quotient spaces, such as the $\rp$ geon
and $\rp$ de Sitter space, where the embedding spaces are Minkowski spaces
with suitable identifications.
We found that the response is related to that of uniformly accelerated detectors given 
in section \ref{sec:quotient}. In particular it is shown in the 
embedding space, and expected on the geon, that the response is not thermal, 
in the sense that it does not satisfy the KMS condition, for most times.
Further it is seen that for a detector switched on in the infinite past 
the response is approximately thermal at early times but does not return to
the thermal response at late times in contrast to expectations
(see e.g \cite{lm:geon,pl:langlois}). If the detector however is turned on at some finite time in the
distant past then the response is approximately thermal
when turned on and returns to being approximately thermal in the distant future.

Lastly we considered some responses on de Sitter space and $\rp$ de Sitter
space.
The regularization of \cite{sc:schlicht} is not easily adaptable to
general motions in these spacetimes, due to the possibility
of spatially closed hypersurfaces in the detector's rest frame.
We argued however for a similar regularization by reinterpreting the
regularization
as an ultraviolet cut-off in the high ``frequency'' modes.
On de Sitter space the transition rate for a detector switched
on in the infinite past is found for a uniformly accelerated 
detector, and it is found to agree with previous literature \cite{dl:deser3}.
This result suggests our regularization is reasonable. In the
case of an inertial detector in de Sitter space the regularization we introduced is shown
to also come from the consideration of a detector with spatial 
extension where the detector is rigid in its rest frame.
On $\rp$ de Sitter space the response of a detector following the
comoving worldline was considered in two ways.
Firstly a direct calculation showed that the response is
exactly that of a uniformly accelerated detector approaching the
boundary on Minkowski space with boundary (identifying $1/\alpha$ with
the acceleration). Again therefore the response of a detector 
switched on in the infinite past has an oscillatory behaviour in the
distant future and does not tend to the expected thermal result, the image
term breaking the KMS condition, in contrast to what was expected \cite{jk:desit}.
The response of a detector switched on at a finite time does
however tend to the expected thermal response at late times.
Secondly we considered the same calculation from the GEMS perspective.
Although the response is clearly different to that in the original space, 
due to the different dimensions of $\rp$ de Sitter space and the embedding
space, it is seen that the response is qualitatively very similar.
The calculation therefore provides a good example for investigating
the relation between detector responses in curved spacetimes and those in their GEMS,
and the use of the GEMS procedure on such quotient spaces where the embedding spaces
are Minkowski spaces with suitable identifications. The calculation further 
suggests that the response found in the GEMS of the $\rp$ geon in section
\ref{sec:Geon} is indeed closely related to the response of the static detector in the
geon itself. 

\subsection*{Acknowledgements}

I would like to thank Jorma Louko for many useful discussions and
for reading the manuscript. This work was supported by the University of
Nottingham and the States of Guernsey Education Department.

\end{document}